\documentclass[conference]{IEEEtran}

\usepackage{amsmath,amssymb}
\usepackage{graphicx}
\usepackage{xcolor}
\usepackage{colortbl}
\usepackage{booktabs}
\usepackage{url}
\usepackage[hidelinks]{hyperref}
\usepackage{orcidlink}
\hypersetup{
  pdftitle={Aneto: Predicting System Performance by Exploiting Cross-Workload Regularity},
  pdfauthor={Raul Taranco, Rene Mueller, Michael Giardino},
  pdfsubject={MICRO 2026},
  pdfkeywords={performance modeling, memory systems, analytical models, performance counters, CPI prediction}
}
\usepackage{balance}
\providecommand{\Description}[1]{}

\usepackage[backend=biber,style=ieee,maxbibnames=99]{biblatex}
\usepackage{tikz}
\usetikzlibrary{decorations.pathmorphing, decorations.pathreplacing, decorations.shapes}
\usetikzlibrary{arrows,arrows.meta}
\usetikzlibrary{plotmarks}
\usetikzlibrary{positioning,fit,backgrounds}
\usepackage{pgfplots}
\pgfplotsset{compat=1.18}

\newcommand{\model}{Aneto}
\usepackage{nicefrac}
\usepackage{multirow}
\usepackage{acro}
\usepackage{microtype}
\usepackage{paralist}

\DeclareAcronym{cpi}{short=CPI, long=cycles per instruction}
\DeclareAcronym{ipc}{short=IPC, long=instructions per cycle}
\DeclareAcronym{mpi}{short=MPI, long=misses per instruction}
\DeclareAcronym{bf}{short=BF, long=blocking factor}
\DeclareAcronym{mlp}{short=MLP, long=memory-level parallelism}
\DeclareAcronym{ilp}{short=ILP, long=instruction-level parallelism}
\DeclareAcronym{mp}{short=MP, long=memory penalty}
\DeclareAcronym{llc}{short=LLC, long=last-level cache}
\DeclareAcronym{rob}{short=ROB, long=reorder buffer}
\DeclareAcronym{ooo}{short=OoO, long=out-of-order}
\DeclareAcronym{hbm}{short=HBM, long=high-bandwidth memory}
\DeclareAcronym{cxl}{short=CXL, long=Compute Express Link}
\DeclareAcronym{nvm}{short=NVM, long=non-volatile memory}
\DeclareAcronym{lpddr}{short=LPDDR, long=low-power double data rate}
\DeclareAcronym{ddr}{short=DDR, long=double data rate}
\DeclareAcronym{dvfs}{short=DVFS, long=dynamic voltage and frequency scaling}
\DeclareAcronym{mape}{short=MAPE, long=mean absolute percentage error}
\DeclareAcronym{ols}{short=OLS, long=ordinary least squares}
\DeclareAcronym{pmu}{short=PMU, long=performance monitoring unit}
\DeclareAcronym{fpga}{short=FPGA, long=field-programmable gate array}
\DeclareAcronym{mshr}{short=MSHR, long=miss status holding register}
\DeclareAcronym{glm}{short=GLM, long=generalized linear model}
\DeclareAcronym{dram}{short=DRAM, long=dynamic random-access memory}
\DeclareAcronym{isa}{short=ISA, long=instruction set architecture}
\DeclareAcronym{loo}{short=LOO, long=leave-one-out}
\DeclareAcronym{ecm}{short=ECM, long=execution-cache-memory}
\DeclareAcronym{carm}{short=CARM, long=cache-aware roofline model}
\DeclareAcronym{dbi}{short=DBI, long=dynamic binary instrumentation}

\newcommand{\resNwlZenTwo}{85}

\newcommand{\resNwlZenThree}{85}

\newcommand{\resNwlZenFour}{70}

\newcommand{\resNwlZenFive}{70}

\newcommand{\resNwlIntel}{78}

\newcommand{\resNwlCsDpc}{130}

\newcommand{\resNwlCsNopf}{129}

\newcommand{\resCpiMedSniper}{14.7}

\newcommand{\resCpiAll}{14.7}
\newcommand{\resNwlAll}{100}
\newcommand{\resNwlMin}{70}
\newcommand{\resNwlMax}{130}

\newcommand{\resAmdBfMedLo}{0.015}
\newcommand{\resAmdBfMedHi}{0.019}
\newcommand{\resAmdCpiMedLo}{14.8}
\newcommand{\resAmdCpiMedHi}{18.2}
\newcommand{\resCsCpiMedLo}{11.6}
\newcommand{\resCsCpiMedHi}{11.7}
\newcommand{\resCsBfMedLo}{0.027}
\newcommand{\resCsBfMedHi}{0.053}
\newcommand{\resIntelCpiPNinety}{27.7}
\newcommand{\resIntelBfMed}{0.050}

\newcommand{\resFmemPNinetyIntel}{0.48}
\newcommand{\resFmemPNinetyZenFour}{0.46}
\newcommand{\resFmemPNinetyZenFive}{0.61}

\newcommand{\resBfRhoCrossZenTwoZenThree}{0.93}

\newcommand{\resBfRhoCrossZenThreeIntel}{0.20}
\newcommand{\resBfNshared}{41}

\newcommand{\resCpiCxldirZenFive}{5.8}

\newcommand{\resBaseNsZenFive}{101}
\newcommand{\resBaseNsIntel}{72}

\newcommand{\resTechPfiftyFirst}{1.5}
\newcommand{\resTechPfiftyLast}{13.5}
\newcommand{\resTechPninetyFirst}{5.7}
\newcommand{\resTechPninetyLast}{39.0}

\newcommand{\resTopNwl}{70}
\newcommand{\resTopNshow}{20}
\newcommand{\resTopPctGood}{57}
\newcommand{\resTopPctModerate}{23}
\newcommand{\resTopPctChallenging}{20}

\newcommand{\resBetaXshared}{0.48}
\newcommand{\resBetaVshared}{0.43}

\newcommand{\resPfXrangeCsDpc}{1.7--1.9}

\newcommand{\resPfPNMedCsDpc}{1.8}
\newcommand{\resPfXrangeCsNopf}{1.1--1.2}

\newcommand{\resPfPNMedCsNopf}{3.9}
\newcommand{\resPfPEightCsDpc}{< 10^{-9}}
\newcommand{\resPfPEightCsNopf}{< 10^{-5}}

\newcommand{\resPfDemPfiftyThree}{12.9}
\newcommand{\resPfAllPfiftyThree}{31.8}
\newcommand{\resPfDemPfiftyEight}{16.9}
\newcommand{\resPfAllPfiftyEight}{43.8}
\newcommand{\resPfBandRange}{23--75}
\newcommand{\resPfCoverRange}{57--63}
\newcommand{\resPfRatio}{1.4}
\newcommand{\resPfPNRatio}{2.8}

\newcommand{\resCalibrationCostZenFive}{8.0}
\newcommand{\resClappMarginalZenFive}{30}
\newcommand{\resAnetoCostZenFive}{5}
\newcommand{\resCostReductionZenFive}{6}

\newcommand{\resCostTrainingWl}{16}

\newcommand{\resLatSensNwl}{70}
\newcommand{\resLatSensNplatforms}{8}

\newcommand{\resLatSensTauMin}{0.70}
\newcommand{\resLatSensTauMax}{0.85}
\newcommand{\resLatSensTauCpiMin}{0.86}
\newcommand{\resLatSensTauCpiMax}{0.92}
\newcommand{\resLatSensTauCpiHiMin}{0.77}
\newcommand{\resLatSensTauCpiHiMax}{0.89}
\newcommand{\resConfCorrect}{59}
\newcommand{\resConfAccuracy}{84}
\newcommand{\resConfErrors}{11}

\newcommand{\resBetaStabDeltaAbsMax}{2.4}
\newcommand{\resBetaStabNplatforms}{8}

\newcommand{\resBetaStabNstarMedian}{5}
\newcommand{\resBetaStabNstarMax}{20}
\newcommand{\resBetaStabNstarMin}{5}

\newcommand{\resBetaStabK}{16}
\newcommand{\resBetaStabMpFactor}{3}
\newcommand{\resBetaStabM}{500}

\newcommand{\resBetaStabDelta}{0.01}

\newcommand{\resMeasClappPZenFive}{0.2}
\newcommand{\resMeasAnetoPZenFive}{1.3}

\newcommand{\resMeasMpRangeZenFive}{0.7}

\newcommand{\resMeasClappPIntel}{0.3}
\newcommand{\resMeasAnetoPIntel}{1.8}

\newcommand{\resMeasMpRangeIntel}{0.4}

\newcommand{\resMeasClappPCsDpc}{1.4}
\newcommand{\resMeasAnetoPCsDpc}{7.5}

\newcommand{\resMeasMpRangeCsDpc}{6.7}

\newcommand{\resMeasMpRangeCsNopf}{7.0}

\newcommand{\resMeasNwlSniper}{89}

\newcommand{\resMeasAnetoPArmFar}{12.7}
\newcommand{\resMeasAnetoPNinetyArmFar}{35.9}

\newcommand{\resMeasMpRangeRArmFar}{3}

\newcommand{\resArmBenefitPctNear}{36}

\DeclareMathOperator{\logit}{logit}

\usepackage{soul}
\sethlcolor{yellow}

\definecolor{hlyellow}{rgb}{1,1,1}

\newcommand{\etal}{\textit{et al.}}

\begin{document}

\title{\model{}: Predicting System Performance by Exploiting Cross-Workload Regularity}

\author{%
\IEEEauthorblockN{Raul Taranco\,\orcidlink{0000-0002-1564-4365}}
\IEEEauthorblockA{\textit{Huawei Technologies}\\
Zurich, Switzerland\\
raul.taranco@h-partners.com}
\and
\IEEEauthorblockN{Rene Mueller\,\orcidlink{0000-0001-6084-9944}}
\IEEEauthorblockA{\textit{Huawei Technologies}\\
Zurich, Switzerland\\
rene.mueller@huawei.com}
\and
\IEEEauthorblockN{Michael Giardino\,\orcidlink{0000-0002-9906-720X}}
\IEEEauthorblockA{\textit{Huawei Technologies}\\
Zurich, Switzerland\\
michael.giardino@huawei.com}
}

\maketitle

\begin{abstract}
Predicting how a workload responds to a change in memory technology requires estimating how much of each cache miss actually stalls the processor. 
Obtaining this stall fraction accurately has traditionally demanded detailed simulation, repeated measurements, or heavy profiling. 
One-shot alternatives exist but sacrifice accuracy. 
We observe that hardware counters from a single native run suffice to infer the stall fraction without simulation. 
Across more than 100 distinct workloads spanning integer, floating-point, graph, and AI benchmarks, the relationship between CPI and the maximum memory stall per instruction follows a predictable pattern on each microarchitecture. 
Aneto is a mechanistic-empirical regression model that exploits this observation. 
Once fitted on a machine across a small set of reference workloads, the model estimates the performance-latency sensitivity of any new workload from a single run, enabling first-order CPI prediction under any memory configuration. 
Aneto reaches \mbox{$1.4\times$} lower median and \mbox{$2.8\times$} lower 90th-percentile CPI error than the best prior one-shot predictor. 
We validate the predictions directly against hardware measurements on an ARM server, from local DDR to HBM and up to \mbox{$\sim$\resMeasMpRangeRArmFar$\times$} the baseline memory penalty, where the median CPI error is \mbox{\resMeasAnetoPArmFar\%} and the 90th percentile \mbox{\resMeasAnetoPNinetyArmFar\%}. At an \mbox{$8\times$} memory-latency extrapolation beyond the reach of direct measurement, Aneto agrees on Zen 5 to within 14.8\% at the median and 42\% at the 90th percentile with a prior method that needs multiple runs per workload. 
Additionally, Aneto provides qualitative insights into workloads and architectures.
\end{abstract}

\section{Introduction}
\label{sec:introduction}

How would workload $X$ perform with twice the bandwidth but 25\% higher latency?
What is the least amount of data needed to answer this question?
Architects evaluating \acs{hbm}, \acs{cxl}, or next-generation DDR need per-workload performance predictions.
However, testing every workload on every candidate memory subsystem across platforms, if even possible, is prohibitively expensive.
In many cases, the memory subsystem, the platform, or both are not physically available, forcing architects to resort to simulation.
Even well-tuned gem5~\cite{Binkert:gem5:2011} simulations can be inaccurate~\cite{Gutierrez:Sources:2014, Akram:Validation:2019}.
Moreover, they run 10{,}000$\times$ slower than native execution, making design-space exploration infeasible.

Analytical models avoid both the cost of measurements on real hardware and the overhead of simulation.
Most analytical \ac{cpi} models separate performance into a \emph{compute} component, governed by the core pipeline and cache hierarchy, and a \emph{memory} component, governed by off-chip access cost.
This decomposition appears in Emma's pipeline analysis~\cite{Emma:Understanding:1997}, Chou~\etal{}~\cite{Chou:MLP:2004}, \ac{cpi} stacks~\cite{Eyerman:CPIComponents:2006}, and the Interval Model~\cite{Eyerman::Mechanistic:2009}.
Clapp~\etal{}~\cite{Clapp:Quantifying:2015} reformulated Chou's~\cite{Chou:MLP:2004} model by introducing a \acf{bf} term.
They express \ac{cpi} as:
\begin{equation}
\mathit{CPI} \;=\; \mathit{CPI}_0 \;+\; \mathit{BF} \cdot \mathit{MPI} \cdot \mathit{MP}
\label{eq:cpi_intro}
\end{equation}
where $\mathit{CPI}$ is the average cycles per instruction, $\mathit{MPI}$ the average \ac{llc} misses per instruction, and $\mathit{MP}$ the miss penalty in cycles per miss.
$\mathit{MPI} \cdot \mathit{MP}$ together represent the memory penalty.
The \acs{bf} quantifies the fraction of the memory penalty that stalls the pipeline, which is reflected in the \acs{cpi}.
This term captures the exploitable \ac{mlp} and the \acs{ooo} overlap~\cite{Chou:MLP:2004}.
If $\mathit{BF} = 0$, the processor overlaps all \acs{llc} misses with useful work, while $\mathit{BF} = 1$ means that every \acs{llc} miss blocks the pipeline for the access' entire duration.
$\mathit{CPI}_0$ is the hypothetical \acs{cpi} of the \emph{compute} component: the \acs{cpi} a workload would exhibit if every \ac{llc} miss were served instantly, without any off-chip memory penalty.
While $\mathit{CPI}$, $\mathit{MPI}$, and $\mathit{MP}$ can be measured directly using standard hardware performance counters, $\mathit{CPI}_0$ and \acs{bf} are not observable and must be inferred.
The \acs{cpi} is the target of the performance prediction.
However, the \acs{bf} is the key parameter that determines the sensitivity of \ac{cpi} to \acs{mp}.

Every existing method for estimating the \acs{bf} incurs a per-workload cost.
\emph{Profiling-based} approaches derive the overlap between miss latency and useful computation from microarchitectural analysis.
For example, the Interval Model~\cite{Eyerman::Mechanistic:2009} analyzes \ac{rob} occupancy, while \ac{cpi} stacks~\cite{Eyerman:CPIComponents:2006} attribute stall cycles to individual miss events.
Both require trace-driven simulation, binary instrumentation, or dedicated counters that commodity hardware does not provide.
Approximate \ac{cpi} decompositions (e.g., Intel Top-Down Microarchitecture Analysis,
IBM POWER CPI accounting~\cite{Eyerman:CPIComponents:2006}) exist, yet none reports the \acs{bf} explicitly.
\emph{Measurement-based} approaches work with standard counters but need multiple runs.
For example, Clapp~\etal{}~\cite{Clapp:Quantifying:2015} run each workload at different memory latencies.
They then fit the linear model from \autoref{eq:cpi_intro} and infer the values of \acs{bf} and $\mathit{CPI}_0$ for the specific workload, memory subsystem and platform.
More broadly, machine-learning models~\cite{Lee:RegressionModeling:2006, Ipek:DSE:2006} and analytical-ML hybrids such as Concorde~\cite{NasrEsfahany:Concorde:2025} can predict performance directly, but produce opaque predictions that resist physical interpretation.
ML approaches usually require vast amounts of training data that can only be generated from simulation, inheriting the simulator's error relative to real hardware.
No existing approach recovers \acs{bf} from a single run.

We find that each platform's fixed resources and their interaction across a diverse set of workloads shape the platform's \acs{bf} response in a predictable way.
The \ac{rob} depth, \ac{mshr} count, and prefetcher determine how much miss latency the hardware can hide behind useful work.
Every workload uses these resources, but differently.
Our key finding is that, despite this diversity, we can approximate the resulting variation in the \acs{bf} with a simple regression using only a small set of standard counters.

We obtain the \ac{bf} through a regression on three standard counters from a single run: \ac{cpi}, \ac{llc} \ac{mpi}, and \ac{mp}.
This predicted \acs{bf} is substituted into \autoref{eq:cpi_intro} to estimate \acs{cpi} under any memory subsystem~(\autoref{sec:model}).

We make three contributions:
\begin{inparaenum}[(1)]
	\item \textbf{Performance prediction of a new workload from a single measurement.}
	We show that a single baseline measurement of three standard counters (\acs{cpi}, \ac{llc} \acs{mpi}, \acs{mp}) suffices to predict workload performance across memory frequencies and technologies (e.g., DDR4 to \ac{hbm}), without per-workload frequency sweeps.
	\item \textbf{Cross-workload predictor of the \ac{bf}.}
	A cross-workload regression model fits the \acs{bf} as a smooth function of log-transformed \acs{cpi} and $\mathit{MPI} \times \mathit{MP}$.
	Across more than \resNwlAll{} workloads evaluated on eight platforms, the model achieves a median absolute \mbox{\acs{cpi}} error of \mbox{\resCpiAll{}\%}.
	On the two ChampSim configurations, the error is \resPfRatio{}$\times$ lower at the median and \resPfPNRatio{}$\times$ lower at the 90th percentile than a state-of-the-art performance model~\cite{Radulovic:PROFET:2019}.
	\item \textbf{Detailed evaluation on \resNwlAll{} workloads, five real platforms, three simulator configurations, across a broad memory latency spectrum.}
	We validate on five physical platforms (AMD Zen2/3/4/5, Intel Comet Lake) for cross-workload generalization, and add an ARM server for direct DDR5-to-\mbox{\acs{hbm}} validation, across seven benchmark suites (more than \resNwlAll{} workloads), with all ground truth from hardware measurements.
	ChampSim simulations, with and without prefetching, together with the Sniper simulator, extend the analysis to memory configurations and prefetch policies that cannot be tested physically, where the model replaces time-consuming re-simulation with a single evaluation.
\end{inparaenum}


\section{Motivating Usage Scenario}
\label{sec:motivation}

Clapp~\etal{}~\cite{Clapp:Quantifying:2015} showed that a workload's performance point is determined by intersecting its bandwidth demand with the memory subsystem's band\-width-la\-ten\-cy curve.
MESS~\cite{Esmaili:MESS:2024} characterizes these curves across technologies.
We illustrate the complete prediction procedure through a concrete scenario: what is the performance of workload $X$ when switching from memory $A$ to $B$ that has twice the bandwidth but 25\% higher latency?

\begin{figure}[t]
	\centering
	\resizebox{\columnwidth}{!}{  
  \begin{tikzpicture}[x=5mm,y=5mm,font=\scriptsize]
  \foreach \y/\l in {1/2, 3.5/1} {
		\node[draw=black, semithick, rotate=90, minimum width=8mm, minimum height=3mm] at (1, \y) {core};
		\node[draw=black, semithick, minimum width=6mm, minimum height=3mm] at (2.07, \y - 0.45) { L1I };
		\node[draw=black, semithick, minimum width=6mm, minimum height=3mm] at (2.07, \y + 0.45) { L1D };
		\node[draw=black, semithick, rotate=90, minimum width=8mm, minimum height=3mm] at (3.18, \y) { L2 };
  }
	\node[draw=black, semithick, rotate=90, minimum width=20.5mm, minimum height=3mm] at (4.15, 2.25) { Last Level Cache };
	\node[rotate=90] at (1, 2.28) { $\cdots$ };
	\node[rotate=90] at (3.18, 2.28) { $\cdots$ };
	\draw[dashed] (0.5,0) rectangle +(4.2,4.48);
	\node[anchor=north] at (2.6, 0) { {\bfseries CPU} };

	\foreach \xshift/\yshift/\c/\label in {10/3.8/black/A, 10/1.2/blue/B} {
		\begin{scope}[shift={(\xshift,\yshift)},\c]  
			\draw[semithick] (0, 0) -- +(3,0);
			\draw[semithick] (0, 1) -- +(3,0);
			\draw[semithick] (3.5,0.5) circle ( 0.5 );
			\foreach \x in {1.0,1.4,...,3} {
				\draw[semithick] (\x, 0) -- +(0, 1);
			}
			\draw[dashed] (-0.25, -0.25) rectangle +(4.5, 1.5);
			\node[anchor=north] at (2.375, -0.25) { {\bfseries Memory \label} };

			\draw[thick,-latex'] (-4.85, 0.5) -- +(4, 0);
			\node[anchor=south, inner sep=1pt] at (-2.85, 0.5) { Memory Access };
			\node[anchor=north, inner sep=1pt] at (-2.85, 0.5) { Demand };

			\draw[thick,-latex'] (4.75, 0.5) -- ++(0.5, 0)
				-- ++(0, -1.5) -- (-4.85, -1);
			\node[anchor=south, inner sep=1pt] at (-2.85, -1) { Supply };
		\end{scope}
	}

  \end{tikzpicture}}
	\Description{CPU-memory model showing the supply side (memory system band\-width-latency response) and demand side (workload memory sensitivity).}
	\caption{CPU-memory model with two memory systems $A$ and $B$. \textit{Supply} characterizes the memory subsystem's band\-width-latency response. \textit{Demand} characterizes each workload's bandwidth requirements.}
	\label{fig:aneto-model}
\end{figure}
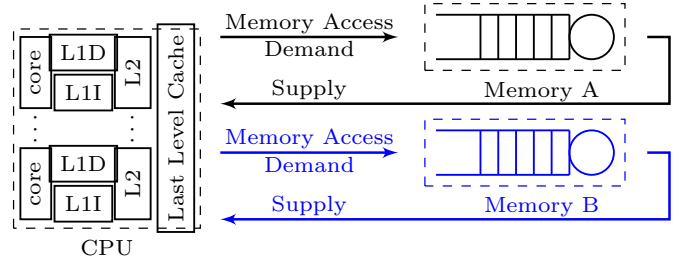

The platform is separated at the \ac{llc} boundary into two components (see \autoref{fig:aneto-model}).
The CPU side (core and caches through the \ac{llc}) determines $\mathit{CPI}_0$, the cost of executing a workload when every \ac{llc} miss is served instantly.
The memory side (controller, interconnect, and chips beyond the \ac{llc}) adds service and queueing latency penalty $\acs{bf} \times \mathit{MPI} \times \mathit{MP}$ from \autoref{eq:cpi_intro}.
This separation enables performance predictions when the memory subsystem changes for a given CPU.
We refer to the two sides as \emph{supply} (memory system response) and \emph{demand} (workload characteristic).

The \emph{supply} side describes the memory system's band\-width-latency response.
MESS~\cite{Esmaili:MESS:2024} measures this response through controlled synthetic benchmarks, producing families of band\-width-latency curves parameterized by read:write ratio, access pattern, and outstanding request count.
For prediction, the supply side reduces to the effective \acf{mp}, measured in core cycles per \ac{llc} miss.

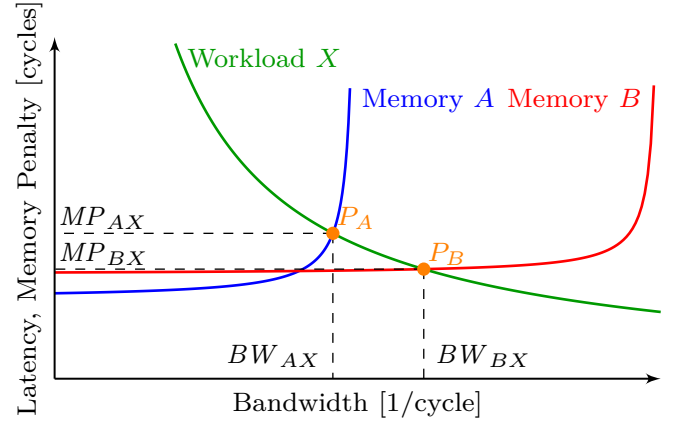
\begin{figure}[t]
	\centering
	\resizebox{\columnwidth}{!}{\begin{tikzpicture}[x=10mm, y=8mm, font=\footnotesize]
  \begin{axis}[
    axis lines=none,
    ticks=none,
    xmin=0, xmax=8,
    ymin=0, ymax=8,
    restrict y to domain=0:8,
    samples=200,
    clip=false,          
    width=80mm,          
    height=52mm,         
    set layers           
    ]
    \addplot[blue, thick, domain=0:3.999] {1/(8 - 2*x) + 1.875};     
    \addplot[red, thick, domain=0:7.990] {1/(20 - 2.5*x) + 2.4375};  
    \addplot[green!60!black, thick, domain=0.5:8] {10 / (0.8*x)};    

    \draw[semithick,-latex'] (0,0) - +(8, 0) node[midway, anchor=north]{Bandwidth [1/cycle]};
    \draw[semithick,-latex'] (0,0) - +(0, 8) node[midway, anchor=south, rotate=90]{Latency, Memory Penalty [cycles]};

    \draw[blue] (3.9, 6.5) node[anchor=west] {Memory $A$};
    \draw[red]  (7.9, 6.5) node[anchor=east] {Memory $B$};
    \draw[green!60!black] (1.6,7.5) node[anchor=west] {Workload $X$};

    \pgfmathsetmacro{\pAx}{3.67285012705684}
    \pgfmathsetmacro{\pAy}{3.4033515029421055}
    \pgfmathsetmacro{\pBx}{4.87253609672063}
    \pgfmathsetmacro{\pBy}{2.565399158030434}

    \draw[dashed] (\pAx, \pAy) -- (\pAx,0) node [anchor=south east] {$\mathit{BW}_{AX}$};
    \draw[dashed] (\pBx, \pBy) -- (\pBx,0) node [anchor=south west] {$\mathit{BW}_{BX}$};

    \draw[dashed] (\pAx, \pAy) -- (0,\pAy) node [anchor=south west, inner sep=1pt] {$\mathit{MP}_{AX}$};
    \draw[dashed] (\pBx, \pBy) -- (0,\pBy) node [anchor=south west, inner sep=1pt] {$\mathit{MP}_{BX}$};

    \fill[orange] (\pAx,\pAy) circle (2pt) node[above right, inner sep=1pt] {$P_A$};
    \fill[orange] (\pBx,\pBy) circle (2pt) node[above right, inner sep=1pt] {$P_B$};

  \end{axis}
\end{tikzpicture}}
	\caption{Supply and demand in band\-width-latency space. Supply curves (blue, red) show the response of memories $A$ and $B$. The demand curve (green) traces workload $X$'s bandwidth demands. Intersections $P_A$ and $P_B$ mark performance points of each memory.}
	\Description{Two supply curves for memory A and B, and a demand curve for workload X, with intersection points P\_A and P\_B marking operating points.}
	\label{fig:supply_demand}
\end{figure}

The \emph{demand} side describes each workload's memory requirements: how many accesses to memory it requests per instruction and how sensitive its performance is to memory latency.
\autoref{fig:supply_demand} shows the supply curves of memory systems $A$ and $B$ and the workload $X$ demand curve, overlaid in the same band\-width-latency graph. While the supply curve can be directly measured, such as in MESS~\cite{Esmaili:MESS:2024}, the demand curve is not directly observable because it depends on \acs{bf}, which must be estimated.

The intersection between the demand and supply curves determines the workload performance.
Switching from memory $A$ to memory $B$ moves this intersection (and thus the performance) from $P_A$ to $P_B$.
In \autoref{fig:supply_demand}, $X$ achieves higher bandwidth at $P_B$ than at $P_A$, despite the 25\,\% higher unloaded latency of memory $B$.
Next, we describe how to predict the performance at $P_B$.

\paragraph{Scenario: Performance prediction of previously unseen workload $X$ on memory $B$}

\begin{table}[t]
\caption{Notation used in the paper.}
\label{tab:notation}
\centering
\small
\begin{tabular}{@{}lp{0.70\columnwidth}@{}}
\toprule
\textbf{Symbol} & \textbf{Description} \\
\midrule
$\mathit{CPI}_{AX}$  & CPI of workload $X$ on memory $A$ \\
$\mathit{CPI}_{0,X}$ & Compute-only CPI of workload $X$ \\
$\mathit{BF}_{AX}$   & Blocking factor of $X$ on memory $A$ \\
$\mathit{MPI}_{AX}$  & \ac{llc} misses per instruction of $X$ on memory $A$ \\
$\mathit{MP}_{AX}$   & Miss penalty on memory $A$ (cycles per \ac{llc} miss) \\
$\mathit{BPI}_{AX}$  & Bytes per instruction of $X$ on memory $A$ \\
$\mathit{BW}_{AX}$   & Demand bandwidth of $X$ on memory $A$ \\
$\mathit{SupplyBW}_B$& Bandwidth-latency curve of memory $B$ \\
$f_\mathrm{mem}$     & Memory CPI fraction: $1 - \nicefrac{\mathit{CPI}_{0,X}}{\mathit{CPI}_{AX}}$ \\
\bottomrule
\end{tabular}
\end{table}

Assume that we have access to a CPU with memory $A$, either real hardware or a simulator.
The goal is to predict performance when the CPU uses memory $B$. The challenge is that we lack access to memory $B$.
Perhaps the memory technology does not exist yet, cannot be paired with that CPU, or time pressure prevents simulation.
However, the supply curves of $B$ \emph{are} available, derived by shifting $A$'s curves to double the bandwidth and increasing the latency by 25\% as shown in \autoref{fig:supply_demand}.

For a deterministic workload with fixed instruction count, performance on $B$ is characterized by
$\mathit{CPI}_{BX}$ (\autoref{tab:notation} lists the notation).
Given a \acs{bf} estimator $\widehat{\mathit{BF}}$ that infers \acs{bf} from observables, $\mathit{CPI}_{BX}$ can be determined from a \emph{single} run of $X$ on system $A$, as follows.

\noindent Step 1: Run $X$ on system $A$. Measure $\mathit{CPI}_{AX}$, $\mathit{MPI}_{AX}$,
$\mathit{MP}_{AX}$, and the number of accesses measured in bytes per instruction $\mathit{BPI}_{AX}$.
Note that $\mathit{BPI}_{AX}$ includes demand loads, write-backs, prefetches, all of which contribute to the memory queueing.

\noindent Step 2: Determine demand bandwidth for $A$.
\[
\mathit{BW}_{AX} \;=\; \frac{\mathit{BPI}_{AX}}{\mathit{CPI}_{AX}}
\]

\noindent Step 3: Find demand bandwidth for $B$.
\[
\mathit{BW}_{BX} = \frac{\mathit{BPI}_{BX}}{\mathit{CPI}_{BX}} =
\frac{\mathit{BPI}_{AX}}{\mathit{CPI}_{0,X} + \mathit{BF}_{AX}\cdot\mathit{MPI}_{AX}\cdot\mathit{MP}_{BX}}
\]
Because the CPU and workload are unchanged, we assert $\mathit{BPI}_{BX} = \mathit{BPI}_{AX}$ and $\mathit{MPI}_{BX}=\mathit{MPI}_{AX}$.
Although both are affected by prefetch timeliness, which in turn is sensitive to the \ac{mp}, we observe no changes over measurement noise. This assumption is also present in other analytical models~\cite{Clapp:Quantifying:2015,Chou:MLP:2004}.

\noindent Step 4: Estimate the blocking factor for $X$. The \acs{bf} is not directly observable and must be inferred from the three counters measured in Step~1.
The underlying analytical model (\autoref{eq:cpi_intro}) treats the \acs{bf} as constant with respect to the \acs{mp} for a given workload~\cite{Clapp:Quantifying:2015}, so $\mathit{BF}_{BX} = \mathit{BF}_{AX}$. Any estimator $\widehat{\mathit{BF}}$ can be used:
\begin{align*}
\mathit{BF}_{AX} &= \widehat{\mathit{BF}}(\mathit{MPI}_{AX}, \mathit{MP}_{AX}, \mathit{CPI}_{AX}) \\
\mathit{BF}_{BX} &= \mathit{BF}_{AX} \,.
\end{align*}

\noindent Step 5: Determine $\mathit{BW}_{BX}$. Writing \autoref{eq:cpi_intro} for both memories and subtracting cancels the unobservable $\mathit{CPI}_{0,X}$, leaving $\mathit{CPI_{BX}}$ in terms of the measured $\mathit{CPI}_{AX}$ and the \ac{mp} difference:
\begin{align*}
\mathit{BW}_{BX} &= \ensuremath{\displaystyle \frac{\mathit{BPI}_{BX}}{\mathit{CPI}_{BX}}} \\
&= \frac{\mathit{BPI}_{AX}}{\mathit{CPI}_{AX} + \mathit{BF}_{AX} \cdot \mathit{MPI}_{AX} \cdot (\mathit{MP}_{BX} - \mathit{MP}_{AX})}
\end{align*}

\noindent Step 6: Find $\mathit{MP}_{BX}$. We solve for the miss penalty at which the workload's bandwidth demand matches the memory system's capacity:
\[
\mathit{BW}_{BX} \;=\; \mathit{SupplyBW}_{B}(\mathit{MP}_{BX})
\]
When the supply curve comes from MESS measurements, the solution can be determined numerically (e.g., using bisection on $\mathit{MP}_{BX}$).
The \ac{cpi} of workload $X$ on memory $B$ is then $\mathit{CPI}_{BX} = \nicefrac{\mathit{BPI}_{AX}}{\mathit{BW}_{BX}}$.

The entire procedure hinges on Step~4, estimating \acs{bf}.
Existing approaches require trace-driven simulation, \ac{dbi} profiling (orders of magnitude slower), or multiple runs at different latencies to obtain the per-workload overlap structure (\autoref{tab:background_comparison}; see \autoref{sec:related}).
\autoref{sec:model} introduces \model{}, a cross-workload \acs{bf} estimator that replaces these costly methods with the single baseline run from Step~1.
It predicts $\mathit{CPI}_{BX}$ given only $\mathit{CPI}_{AX}$, $\mathit{MPI}_{AX}$, and $\mathit{MP}_{AX}$.

\begin{table}[t]
\caption{State-of-the-art analytical performance prediction}
\label{tab:background_comparison}
\small
\begin{tabular*}{\columnwidth}{@{\extracolsep{\fill}}llc@{}}
\toprule
\textbf{Model} & \textbf{Method} & \textbf{Per workload} \\
\midrule
Interval~\cite{Eyerman::Mechanistic:2009}         & simulator   & 1 trace      \\
Van den Steen~\cite{VanDenSteen:MicroarchitectureIndependent:2016} & DBI profiler & 1 profile \\
Ipek~\cite{Ipek:DSE:2006}                         & simulator & sim.\ data     \\
Clapp~\cite{Clapp:Simple:2015}                     & HW/sim.  & $N$ runs \\
PROFET~\cite{Radulovic:PROFET:2019}                & HW/sim.    & 1 run + BF bounds \\
\textbf{\model{} (ours)}                           & \textbf{HW/sim.} & \textbf{1 run}    \\
\bottomrule
\end{tabular*}
\end{table}

\section{\model{} Model}
\label{sec:model}

The prediction pipeline from \autoref{sec:motivation} reduces to one unknown: the blocking factor.
Predicting the \acs{bf} from a single run would eliminate the per-workload cost of existing methods, but only if the \acs{bf} is predictable across workloads.
We find that in fact it is.

When visualized in log-transformed $(\mathit{MPI} \cdot \mathit{MP},\; \mathit{CPI})$ space, the \acs{bf} varies smoothly across more than 100 diverse workloads, forming a regularity that a simple regression model can capture.
\model{} exploits this regularity.
Because the method requires no frequency scaling or memory throttling, it works equally well on simulators, where Clapp's approach would require re-simulating at each latency point.

\subsection{Empirical Regularity}
\label{sec:model:regularity}

\begin{figure}[t]
	\centering
	\includegraphics[width=\columnwidth]{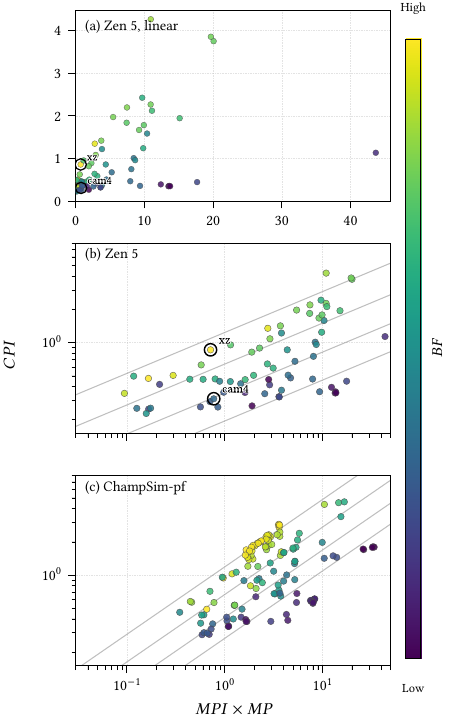}
	\caption{%
		Workloads in the $(\mathit{MPI} \cdot \mathit{MP},\; \mathit{CPI})$ plane, colored by \acs{bf} on a logarithmic color scale normalized per panel.
		\textbf{(a)}~Linear axes show no structure;
		\textbf{(b,\,c)}~log axes reveal a smooth \acs{bf} gradient on Zen\,5 and ChampSim-pf.
		\texttt{cam4} and \texttt{xz} (circled) share $\mathit{MPI} \cdot \mathit{MP}$ but differ $15\times$ in \acs{bf}.}
	\Description{Three scatter plots stacked vertically: linear axes showing no structure, then log-log Zen 5 and ChampSim-pf each showing a smooth diagonal BF gradient.}
	\label{fig:bf_gradient}
\end{figure}

Two workloads can have identical memory exposure $\mathit{MPI} \cdot \mathit{MP}$ yet radically different blocking factors.
Consider the \texttt{cam4}/\texttt{xz} pair (circled in \autoref{fig:bf_gradient}b).
Both sit at $\mathit{MPI} \cdot \mathit{MP} \approx 0.7$, yet \texttt{cam4} has $\acs{bf} = 0.026$ while \texttt{xz} has $\acs{bf} = 0.39$.
The memory exposure alone cannot distinguish them.
What additional information separates a workload that hides latency from one that stalls?

\autoref{fig:bf_gradient}a plots $\mathit{MPI} \cdot \mathit{MP}$ against \ac{cpi} for all qualifying workloads (see \autoref{sec:methodology:benchmarks}) on Zen\,5, coloring each point by $\log(\acs{bf})$.
In linear coordinate axes, the scatter plot appears featureless. Switching to logarithmic axes (\autoref{fig:bf_gradient}b) reveals a structure.
The \acs{bf} varies monotonically across the $(\log(\mathit{MPI} \cdot \mathit{MP}),\; \log(\mathit{CPI}))$ plane so that workloads that have the same \acs{bf} appear on the same gray line, i.e.,
straight iso-contour lines and perpendicular gradients.

What is surprising is not the trend itself, but how regular it is: three parameters in log-space capture the structure across every platform's workload corpus, drawn from SPEC, PARSEC, GAP, and other suites.
\autoref{fig:bf_gradient}c shows that the same regularity also appears on ChampSim and Comet Lake (not shown).
To our knowledge, no prior work has exploited this cross-workload regularity to estimate the \acs{bf}, and thus performance, from a single run.

The answer to the \texttt{cam4}/\texttt{xz} puzzle is the \ac{cpi} itself.
A high $\mathit{MPI} \cdot \mathit{MP}$ paired with a low \ac{cpi} means that the latency is hidden effectively, implying a low \acs{bf}.
The same memory exposure paired with a high \ac{cpi} indicates pipeline stalls, implying a high \acs{bf}.
The \ac{cpi} separates workloads that share memory pressure but differ in compute character.

\subsection{Cross-Workload Estimator}
\label{sec:model:estimator}

We present the model in two steps. We first derive the model and then apply two minor technical improvements.

We estimate the \mbox{\acs{bf}} from the \mbox{\ac{cpi}}, \mbox{\ac{mpi}}, and \mbox{\ac{mp}}.
To capture the empirical regularity described in the previous section, we leverage a linear regression in logarithmic space.
The linear model for the \acs{bf} is
\begin{align}
	\log(\acs{bf}) &= \beta_0 + \beta_1 \log(\mathit{MPI} \cdot \mathit{MP}) + \beta_2 \log(\mathit{CPI}). 	\label{eq:log_bf}
\end{align}
%

For a fixed \acs{bf}, the right-hand side of \autoref{eq:log_bf} is a line in the logarithmic coordinates, which matches the straight lines of \autoref{fig:bf_gradient}.
This is the core of the model. We apply two practical refinements that ensure the \acs{bf} estimates are correctly bounded by construction.

Ignoring the compute component (\mbox{$\mathit{CPI}_0$}) in \autoref{eq:cpi_intro} yields the upper bound \mbox{$\acs{bf}_{\mathrm{ub}} = \mathit{CPI}/(\mathit{MPI} \cdot \mathit{MP})$}, the \mbox{\acs{bf}} at which the memory term alone accounts for the \mbox{\ac{cpi}}.
A \mbox{\acs{bf}} above \mbox{$\acs{bf}_{\mathrm{ub}}$} would imply a negative \mbox{$\mathit{CPI}_0$}.
We normalize the \acs{bf} by $\acs{bf}_{\mathrm{ub}}$
\begin{align}
	\log(\nicefrac{\acs{bf}}{\acs{bf}_{\mathrm{ub}}}) = \beta_0 &+ (\beta_1 + 1) \log(\mathit{MPI} \cdot \mathit{MP}) \nonumber \\
	                                                            &+ (\beta_2 - 1) \log(\mathit{CPI}). \nonumber
\end{align}
%

This is just a different representation of the model from \autoref{eq:log_bf}.
The normalized $\nicefrac{\acs{bf}}{\acs{bf}_{\mathrm{ub}}}$ lies between zero and one.
We take the logit of this as the regression target, since the logit maps \mbox{$(0, 1)$} onto the real line.
The inverse logit keeps every predicted ratio inside \mbox{$(0, 1)$}, so the predicted \mbox{\acs{bf}} stays below \mbox{$\acs{bf}_{\mathrm{ub}}$} and the predicted \mbox{$\mathit{CPI}_0$} positive.
The final model is
\begin{align}
	\logit(\nicefrac{\acs{bf}}{\acs{bf}_{\mathrm{ub}}}) = \beta_0 &+ \beta_1 \log(\mathit{MPI} \cdot \mathit{MP}) \nonumber \\
	                                                              &+ \beta_2 \log(\mathit{CPI}) \label{eq:logit_bf}
\end{align}
where $\logit(p) = \log\big(p/(1-p)\big)$.

We fit the three coefficients by ordinary least squares.\footnote{Least squares is the maximum-likelihood estimator when the residuals are normally distributed. Otherwise a generalized linear model with an explicit variance model could improve estimation efficiency. Least squares performs well on the metric we report, the held-out \acs{cpi} error (\autoref{sec:exp}).}
The ground-truth \mbox{\acs{bf}} in the target comes from Clapp's multi-point regression.
The $\mathit{MP}$ in this equation is the observed baseline penalty, a fixed property of the system.

The regression model exploits \emph{cross-workload} variation, the different memory exposures that different workloads produce at the same operating point.

The two features are necessary because \acs{cpi} conflates two latent quantities.
$\log(\mathit{MPI} \cdot \mathit{MP})$ captures the maximum potential memory stall, while $\log(\mathit{CPI})$ disambiguates. Conditioned on the same memory exposure, the \acs{cpi} variation isolates the blocking factor (\autoref{sec:model:regularity}).
The \texttt{cam4}/\texttt{xz} pair makes this concrete.
Decomposing the \ac{cpi} via \autoref{eq:cpi_intro} shows that the $\mathit{CPI}_0$ accounts for 94\% of \texttt{cam4}'s \ac{cpi} ($0.29$ of $0.31$) but only 68\% of \texttt{xz}'s ($0.58$ of $0.86$).
Exposure alone would entirely miss this difference in behavior.
The estimator resolves a $15\times$ \acs{bf} difference that $\log(\mathit{MPI} \cdot \mathit{MP})$ alone cannot.

Both Clapp's approach and \model{} infer the \acs{bf} from observables but they differ in the identification strategy.
Clapp's approach varies \ac{mp} across $N$ operating points and observes how \ac{cpi} changes (within-workload variation).
\model{} holds the \ac{mp} fixed and exploits the cross-workload distribution (between-workload variation) to estimate the mapping.
\Ac{loo} cross-validation (\autoref{sec:exp}) confirms that the model generalizes to held-out workloads.
\autoref{fig:bf_gradient}b shows this directly. Each point is one workload at a single \ac{mp}, and the smooth color gradient means that the \acs{bf} can be read from position alone, without tracing how the point moves as the \ac{mp} changes.

In our experiments (omitted for space), we confirmed that \autoref{eq:logit_bf} is the most parsimonious form.
Dropping either feature degrades accuracy, and adding an interaction term does not reduce \ac{loo} error consistently across platforms.

%
%
%
%
\subsection{Prediction Pipeline}
\label{sec:model:pipeline}

Applying the logistic function to \autoref{eq:logit_bf} and rescaling by $\acs{bf}_{\mathrm{ub}}$ yield the closed-form estimator
\begin{equation}
	\widehat{\acs{bf}} \;=\; \acs{bf}_{\mathrm{ub}} \cdot \frac{e^{\beta_0} \cdot (\mathit{MPI} \cdot \mathit{MP})^{\beta_1} \cdot \mathit{CPI}^{\beta_2}}{1 + e^{\beta_0} \cdot (\mathit{MPI} \cdot \mathit{MP})^{\beta_1} \cdot \mathit{CPI}^{\beta_2}}.
	\label{eq:bf_closed}
\end{equation}
Any method can supply the \acs{bf} estimate that the procedure from \autoref{sec:motivation} requires in Step~4, whether Clapp's multi-point regression, PROFET's free-parameter band~\cite{Radulovic:PROFET:2019}, or platform-specific sources.
\model{} uses \autoref{eq:logit_bf}, whose advantage is single-run cost with no hardware reconfiguration.

Given the single baseline run of workload $X$ on system $A$ ($\mathit{CPI}_{AX}$, $\mathit{MPI}_{AX}$, $\mathit{MP}_{AX}$), the predicted \ac{cpi} on system $B$ follows directly
\begin{equation}
	\mathit{CPI}_{BX} \;=\; \mathit{CPI}_{AX} + \widehat{\acs{bf}} \cdot \mathit{MPI}_{AX} \cdot (\mathit{MP}_{BX} - \mathit{MP}_{AX}).
	\label{eq:cpi_pred}
\end{equation}
Taking the difference of \acs{mp} between $A$ and $B$ avoids determining $\mathit{CPI}_{0,X}$.
What remains is the \ac{mp} difference $(\mathit{MP}_{BX} - \mathit{MP}_{AX})$, which is the architect's dependent variable.


When $\mathit{MP}_{BX}$ is not known a priori (e.g., because it depends on the load-dependent bandwidth), the supply-demand intersection from \autoref{sec:motivation} (Step~6) can be determined
numerically, e.g., via the bisection method.

The memory fraction
\begin{equation}
	f_{\mathrm{mem}} \;=\; \frac{\acs{bf} \cdot \mathit{MPI} \cdot \mathit{MP}}{\mathit{CPI}} \;=\; \frac{\acs{bf}}{\acs{bf}_{\mathrm{ub}}} \;=\; 1 - \frac{\mathit{CPI}_0}{\mathit{CPI}}
	\label{eq:fmem}
\end{equation}
controls how much of a \acs{bf} error reaches the \ac{cpi}: a 10\% \acs{bf} error at $f_{\mathrm{mem}} = 5\%$ produces a 0.5\% \ac{cpi} error, while the same error at $f_{\mathrm{mem}} = 50\%$ produces 5\%.

\subsection{Scope and Assumptions}
\label{sec:model:scope}

The prediction pipeline inherits two properties from the linear \ac{cpi} decomposition of Clapp~\etal{}~\cite{Clapp:Quantifying:2015} and Chou~\etal{}~\cite{Chou:MLP:2004} that bound its applicability.
The first is that \acs{bf} and $\mathit{CPI}_0$ remain constant when \ac{mp} varies.
\autoref{eq:cpi_intro} confines all latency dependence to the memory term, so $\mathit{CPI}_0$ is independent of memory timing by construction.
The second is that \ac{mpi} does not change with \ac{mp}, because \ac{llc} misses depend on the cache hierarchy and the workload's access pattern, not on \ac{dram} timing.
Prefetcher-sensitive workloads are the natural concern: slower memory increases bandwidth pressure, which can reduce prefetch timeliness and shift the effective \ac{mpi}.
In practice, we did not observe significant \ac{mpi} variability across either the \ac{dvfs} sweep on hardware or the latency sweep in ChampSim, even for workloads with aggressive prefetching.
\autoref{sec:methodology} describes the experimental methodology we use to extract ground-truth \acs{bf} and $\mathit{CPI}_0$ values, which in turn revalidate the linear \ac{cpi} model on modern hardware.

\section{Experimental Methodology}
\label{sec:methodology}

This section describes how we obtain ground-truth \acs{bf} values, defines the validation protocol, and details the platforms, benchmarks, and metrics used in \autoref{sec:exp}.

\subsection{Evaluation Platforms}

We evaluate \mbox{\model{}} across the six hardware platforms listed in \mbox{\autoref{tab:platforms}}, alongside two simulators (ChampSim and Sniper~\cite{Carlson:Sniper:2011}). The five Zen and Intel machines and these three simulator configurations form the eight-platform cross-workload corpus, leaving the sixth machine, an ARM server, for the worked example.
The physical hardware spans four AMD generations (Zen\,2 through Zen\,5), an Intel Comet Lake processor, and an ARM server.
On ChampSim~\mbox{\cite{Gober:ChampSim:2022}}, we run the DPC-4\footnote{4th Data Prefetching Championship, co-located with HPCA 2026.} traces at full \mbox{\acs{dram}} bandwidth, with the default prefetchers enabled (ChampSim-pf, Pythia+\allowbreak Berti) and disabled (ChampSim-nopf), which together expose a wider range of effective \mbox{\acs{mpi}} than either alone.

\begin{table}[t]
  \centering
  \caption{Evaluation platforms configurations and qualifying workloads ($W$) after discarding.}
  \label{tab:platforms}
  \small
  \begin{tabular*}{\columnwidth}{@{\extracolsep{\fill}}lccr@{}}
    \toprule
    Platform & Memory & Freq.\ (GHz) & $W$ \\
    \midrule
    Zen\,2 (Ryzen 9 3950X)       & DDR4-2133 & 2.20--3.20 & \resNwlZenTwo \\
    Zen\,3 (EPYC 7543)           & DDR4-3200 & 1.50--2.80 & \resNwlZenThree \\
    Zen\,4 (Threadripper 7980X)  & DDR5-4800 & 2.20--3.20 & \resNwlZenFour \\
    Zen\,5 (Ryzen 9 9950X3D)     & DDR5-4800 & 2.98--4.30 & \resNwlZenFive \\
    Comet Lake (Xeon W-1270)     & DDR4-2933 & 1.50--3.40 & \resNwlIntel \\
    ARM Server            & DDR5-5600   & 1.20--2.00 & 70 \\
                          & HBM (near)  & 1.20--2.00 & 70 \\
                          & HBM (far)   & 1.20--2.00 & 70 \\
    \midrule
    ChampSim-pf                  & simulated & 4.00       & \resNwlCsDpc \\
    ChampSim-nopf                & simulated & 4.00       & \resNwlCsNopf \\
    Sniper                & simulated & 2.66       & \resMeasNwlSniper \\
    \bottomrule
  \end{tabular*}
\end{table}

\begin{table*}[t!]
  \centering
  \caption{Benchmark suites and workload composition. Graph suites use LJ, Pokec, Kron, URAND, road, and web graphs.}
  \label{tab:workloads}
  \small
  \begin{tabular}{lrl}
    \toprule
    Suite & $W$ & Description \\
    \midrule
    SPEC CPU2017~\cite{Bucek:SPEC2017:2018} & 36 & Speed benchmarks (intrate + fprate), ref inputs run to completion \\
    PARSEC~\cite{Bienia:PARSEC:2008}        & 10 & Parallel kernels (blackscholes, canneal, streamcluster, dedup, \ldots) \\
    GAP~\cite{Beamer:GAP:2015}              & 30 & 6 graph algorithms (BFS, BC, PR, CC, SSSP, TC) $\times$ 5 input graphs \\
    Ligra~\cite{Shun:Ligra:2013}            & 10 & Graph processing (BFS, PageRank, BellmanFord, Triangle, \ldots) on LJ \\
    GMS~\cite{Besta:GMS:2021}               &  5 & Graph mining (clique, coloring, triangle counting) on Pokec; x86-64 only \\
    DaCapo~\cite{Blackburn:DaCapo:2006}     & 10 & JVM applications (h2, batik, jython, avrora, \ldots) \\
    AI-ML                                   &  9 & AI inference: Whisper, Qwen2, TinyLLaMA, ViT, CLIP, Stable Diffusion, BioGPT, RWKV, Bark \\
    \midrule
    DPC-4 traces                            & 132 & 4th Data Prefetching Championship traces, run on ChampSim and Sniper \\
    \bottomrule
  \end{tabular}
\end{table*}

\subsection{Benchmark Suites}
\label{sec:methodology:benchmarks}

\autoref{tab:workloads} lists the seven benchmark suites and the DPC-4 traces, totaling 110~workloads on hardware and 132~on ChampSim.
The suites span general-purpose integer and floating-point codes (SPEC CPU2017), parallel kernels (PARSEC), and irregular pointer-chasing with high \acs{bf} (GAP, Ligra, GMS).
We also include DaCapo (JVM workloads) and a custom AI-ML suite of representative models and CPU inference engines.
We apply two criteria to fit \mbox{\model{}}. We omit workloads whose \mbox{\acs{bf}} fit has an \ensuremath{\displaystyle R^2 < 0.90} (see \autoref{sec:methodology:gt}).
We also discard a small number of workloads that produce parameters outside their valid range (\mbox{\acs{bf}}~\ensuremath{\displaystyle > 1} or \ensuremath{\displaystyle \mathit{CPI}_0 < 0}).
In our experiments, these discarded workloads issue too few misses for any latency sensitivity to register.

\mbox{\model{}} can predict any workload. But when we report the aggregate error, we leave out the memory-inactive ones, those with an \mbox{\acs{mpi}} below a threshold (\ensuremath{\displaystyle <10^{-4}}). Their \mbox{\acs{cpi}} \ensuremath{\displaystyle \approx\mathit{CPI}_0}, so any method predicts them almost perfectly. Keeping them in would only benefit any method.
The $W$ column in \autoref{tab:platforms} reports the qualifying workload count per platform after discarding.
We run each workload independently in a single-thread configuration and bind it onto an isolated core.

\subsection{Measurement and Ground-Truth Estimation}
\label{sec:methodology:gt}

The \ac{bf} is not directly observable from hardware counters.
We estimate a ground-truth \ac{bf} for each workload by measuring \ac{cpi}, \ac{mpi}, and \ac{mp} at $N{=}5$ operating points and regressing the \ac{cpi} against the $\ac{mpi} \times \ac{mp}$ product, following Clapp~\etal{}~\cite{Clapp:Simple:2015}.
The slope of this per-workload fit yields the \ac{bf}, and the intercept yields the $\mathit{CPI}_0$.
We use five points, which suffice to fit both parameters and still detect poor fits via $R^2$.

We call the platform's native operating point the \emph{baseline}, the nominal core frequency on hardware (and thus the nominal \ac{mp} in core cycles) or the default unloaded \ac{dram} latency on ChampSim.
We collect \ac{cpi}, \ac{mpi}, and \ac{mp} at this point to form a \emph{baseline measurement} and predict the \ac{cpi} at non-baseline operating points for different \ac{mp} values.

On hardware, we vary the frequency of the CPU core across $N{=}5$ equally spaced values within the range supported on each platform (\autoref{tab:platforms}).
This spans the widest \ac{mp} variation the hardware allows.
Lowering the core frequency reduces the number of core cycles per unit of memory latency (because \ac{dram} timing remains fixed in nanoseconds while the core clock period grows), which reduces the \ac{mp} measured in core cycles.
At each operating point, we lock core and uncore frequencies and disable turbo boost to eliminate dynamic frequency scaling.
The measurement core is isolated from OS scheduling, reducing run-to-run \ac{cpi} variation to below 1\%.
Each workload runs five independent invocations per operating point, each executing from start to finish, and we report the median.
On ChampSim and Sniper~\mbox{\cite{Carlson:Sniper:2011}}, we modify the \ac{dram} timing parameters to simulate seven memory latencies from $1\times$ to $8\times$ the baseline and vary \ac{mp} directly.

\begin{figure}[t]
  \centering
  \includegraphics[width=\columnwidth]{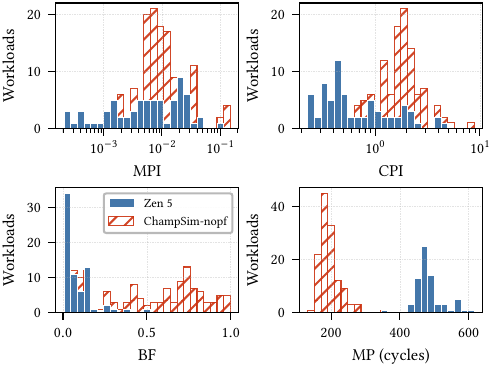}
  \caption{Metric distributions for Zen\,5 and ChampSim-nopf after discarding workloads with $R^2 < 0.90$. Without prefetching, the \ac{bf} distribution shifts toward 1.0.}
  \label{fig:metric_histograms}
\end{figure}

\autoref{fig:metric_histograms} shows the distribution of key metrics across the evaluation corpus.
On Zen\,5, the \ac{bf} clusters near zero where hardware prefetching hides most memory latency.
On ChampSim-nopf, the \ac{bf} extends to ${\approx}1.0$ where nearly every miss stalls the pipeline.

\subsection{Evaluation Protocol}
\label{sec:methodology:eval}

\subsubsection{Cross-Validation}

All results use \acf{loo} cross-validation with a two-step procedure.
First, for each of the $W$ qualifying workloads on a platform, the \acs{bf} regression is fitted by ordinary least squares on the remaining $W{-}1$ workloads.
The held-out workload's \acs{bf} is then predicted from its single baseline measurement.
Second, the predicted \acs{bf} is substituted into the \ac{cpi} equation (\autoref{eq:cpi_pred}) to extrapolate the \ac{cpi} at the target \ac{mp}.
With $W = \resNwlMin\text{--}\resNwlMax$ and three regression parameters, the \ac{loo} procedure completes in under one second per platform and provides a nearly unbiased error estimate.

\subsubsection{Extrapolation Targets and Ground Truth}

We evaluate at $3\times$ the baseline \ac{mp}, representative of \ac{cxl}-attached memory (${\approx}240$\,ns from an ${\approx}80$\,ns DDR5 baseline).
To stress-test the limits of load-store semantics, we also evaluate at $8\times$, representing cross-rack fabric latencies (${\approx}640$\,ns).
The \ac{dvfs} sweep covers only $1.4$--$2.5\times$, so both targets lie well beyond the \ac{mp} range we observe during the ground-truth fit on hardware platforms.

Measuring \ac{cpi} at these \ac{mp} factors on real hardware is impractical, which is precisely the scenario that motivates a predictive model like \model{}.
On hardware, we evaluate how well \model{} approximates the Clapp~\etal{} per-workload extrapolation (\autoref{sec:methodology:gt}).
Within the \ac{dvfs} range, errors reflect deviation from real measurements.
Beyond this range, we have no real measurements, so both \model{} and the Clapp~\etal{} extrapolation rely on the same linear \ac{cpi} model.
Reported errors at $3\times$ or $8\times$ \ac{mp} therefore represent model-vs-model agreement.
On ChampSim, ground-truth \ac{cpi} at any \ac{mp} factor comes from direct simulation.

\subsubsection{Metrics and Statistical Tests}

We employ several metrics to evaluate our model and compare against prior art.
The \emph{relative \ac{cpi} error} $|\mathit{CPI}_{\mathrm{pred}} - \mathit{CPI}_{\mathrm{true}}| / \mathit{CPI}_{\mathrm{true}}$ measures prediction error at the target \ac{mp}.
It is reported at the $50^{\text{th}}$ and $90^{\text{th}}$ percentiles across workloads.
The \emph{absolute \acs{bf} error} $|\Delta\text{BF}|$ isolates the quality of the \ac{bf} prediction from the amplification effect of the $\mathit{MPI} \times \mathit{MP}$ product on the \ac{cpi}.

An architect comparing memory technologies needs to know which workloads suffer most, not just by how much.
We rank workloads by their \ac{cpi} degradation at the target \ac{mp} and compute \emph{Kendall's $\tau$} between the ground-truth and predicted rankings.
A $\tau$ near 1.0 means the model preserves the relative ordering of workloads by memory sensitivity.

When comparing \model{} against alternative methods such as PROFET~\cite{Radulovic:PROFET:2019} (\autoref{sec:exp:profet}), we pair each workload's absolute \ac{cpi} error under both methods at the same \ac{mp} factor and apply the one-sided Wilcoxon signed-rank test ($\alpha = 0.05$).
This non-parametric test determines whether one method produces systematically lower errors than the other without assuming normality.
We apply Holm--Bonferroni correction when testing across multiple platforms.

\section{Evaluation}
\label{sec:exp}

We test \model{} on the \resLatSensNplatforms{} platforms and workload corpus described in \autoref{sec:methodology}, using the \ac{loo} protocol from \autoref{sec:methodology:eval} throughout.
We first assess accuracy against ground truth (\autoref{sec:exp:measurement}), then assess extrapolation accuracy (\autoref{sec:exp:accuracy}) and decision quality (\autoref{sec:exp:decisions}).
To look beyond the accuracy of the model, we examine whether the regression coefficients transfer across processor generations (\autoref{sec:exp:insight}).
We identify limits of the model and its costs (\autoref{sec:exp:limits}). \autoref{sec:exp:profet} closes by comparing against PROFET~\cite{Radulovic:PROFET:2019}, the state-of-the-art single-measurement predictor.

\subsection{Accuracy Against Ground Truth}
\label{sec:exp:measurement}

\autoref{tab:measurement} compares both the per-workload fit of Clapp~\etal{}~\cite{Clapp:Quantifying:2015} and \model{} against the \ac{cpi} measured at each operating point.
Each workload has an \ac{mp} factor which is calculated by dividing the measured \ac{mp} that is most distant from the baseline \ac{mp} by the baseline \ac{mp}. This provides the relative measurable \ac{mp} range in our experiments.
The \ac{mp} factor column is the median \ac{mp} factor for all workloads on each platform.
On hardware, the \ac{dvfs} range reduces the \ac{mp} to \resMeasMpRangeZenFive$\times$--\resMeasMpRangeIntel$\times$ of the baseline, while on ChampSim, the latency range increases it to \resMeasMpRangeCsDpc$\times$--\resMeasMpRangeCsNopf$\times$.

To test the linearity assumption underlying Clapp~\etal{}~\cite{Clapp:Quantifying:2015}, we apply a leave-one-point-out test: for each workload, we fit the linear model on all but one operating point and predict the \ac{cpi} at the held-out point.
The resulting P50 (median) error is \resMeasClappPZenFive\% on Zen\,5 and at most \resMeasClappPIntel\% on Comet Lake, confirming that the linear \ac{cpi} decomposition is accurate to at most \resMeasClappPIntel\% on all five hardware platforms.
On ChampSim, where the \ac{mp} range reaches \resMeasMpRangeCsDpc$\times$ from the baseline, the error increases to \resMeasClappPCsDpc\% (ChampSim-pf).
Even with this variation in \ac{mp}, the linear model remains accurate to at most \resMeasClappPCsDpc\%, providing a sound basis for extrapolation beyond the measured range.

\model{} replaces the per-workload frequency sweep with a single baseline measurement and predicts the \ac{cpi} at each non-baseline operating point via the \ac{loo} protocol.
\model{} achieves a P50 error of \resMeasAnetoPZenFive\% on Zen\,5 and \resMeasAnetoPIntel\% on Comet Lake.
Our method avoids the multi-point sweep at the cost of a modest increase over Clapp~\etal{}~\cite{Clapp:Quantifying:2015}.
On ChampSim, the error increases to \resMeasAnetoPCsDpc\% P50 (ChampSim-pf), due to the high extrapolation range and the inherited error from Clapp~\etal{}~\cite{Clapp:Quantifying:2015} used as ground-truth in \model{}.

\mbox{\autoref{tab:measurement}} groups this comparison by how far each regime can push the penalty against direct ground truth. On hardware the CPU frequency can only vary the memory penalty over a small range, limiting the validation scale. The simulators reach up to \ensuremath{\displaystyle 7\times} the baseline penalty against exact ground truth.

\begin{table}[t]
  \centering
  \caption{Accuracy against ground-truth \ac{cpi}, grouped by regime: the hardware frequency sweep and the simulators.}
  \label{tab:measurement}
  \small
  \begin{tabular}{@{}cl r rr rr@{}}
    \toprule
    & & & \multicolumn{4}{c@{}}{Relative CPI error (\%)} \\
    & & & \multicolumn{2}{c}{Clapp~\etal{}} & \multicolumn{2}{c@{}}{\model{}} \\
    \cmidrule(lr){4-5} \cmidrule(l){6-7}
    & Platform & \ac{mp} factor & P50 & P90 & P50 & P90 \\
    \midrule
    \multirow{5}{*}{\rotatebox[origin=c]{90}{\small Freq.\ sweep}} & Zen\,2 & 0.7$\times$ & 0.1 & 0.7 & 0.9 & 4.4 \\
     & Zen\,3 & 0.5$\times$ & 0.1 & 0.6 & 1.3 & 5.4 \\
     & Zen\,4 & 0.7$\times$ & 0.1 & 0.6 & 1.1 & 4.1 \\
     & Zen\,5 & 0.7$\times$ & 0.2 & 1.1 & 1.3 & 5.3 \\
     & Comet Lake & 0.4$\times$ & 0.3 & 1.8 & 1.8 & 6.2 \\
    \midrule
    \multirow{3}{*}{\rotatebox[origin=c]{90}{\small Sim.}} & ChampSim-pf & 6.7$\times$ & 1.4 & 5.6 & 7.5 & 28.5 \\
     & ChampSim-nopf & 7.0$\times$ & 0.8 & 2.7 & 7.7 & 21.3 \\
     & \cellcolor{hlyellow}Sniper & \cellcolor{hlyellow}5.8$\times$ & \cellcolor{hlyellow}0.1 & \cellcolor{hlyellow}0.5 & \cellcolor{hlyellow}8.0 & \cellcolor{hlyellow}30.7 \\
    \bottomrule
  \end{tabular}
\end{table}

\subsection{Extrapolation Accuracy}
\label{sec:exp:accuracy}

\autoref{sec:exp:measurement} established that within the measured range, \model{} introduces a P50 error of \resMeasAnetoPZenFive\% on Zen\,5.
We now stress-test by extrapolating beyond that range.
\autoref{tab:accuracy} reports the \ac{cpi} prediction error and the absolute \ac{bf} error at $8\times$ the baseline unloaded memory latency (${\approx}640$\,ns from an 80\,ns DDR5 baseline).

\begin{table}[t]
  \centering
  \caption{Prediction accuracy under \ac{loo}. CPI error (\%): P50 and P90 at $8\times$ \ac{mp}. $|\Delta\text{BF}|$: absolute \ac{bf} prediction error.}
  \label{tab:accuracy}
  \small
  \begin{tabular}{@{}cl rr rr@{}}
    \toprule
    & & \multicolumn{2}{c}{CPI error (\%)} & \multicolumn{2}{c@{}}{$|\Delta\text{BF}|$} \\
    \cmidrule(lr){3-4} \cmidrule(l){5-6}
    & Platform & P50$_{8\times}$ & P90$_{8\times}$ & P50 & P90 \\
    \midrule
    \multirow{5}{*}{\rotatebox[origin=c]{90}{\small Freq.\ sweep}} & Zen\,2       & 18.2 & 52.0 & 0.016 & 0.058 \\
     & Zen\,3       & 16.5 & 46.9 & 0.019 & 0.094 \\
     & Zen\,4       & 17.3 & 53.3 & 0.019 & 0.086 \\
     & Zen\,5       & 14.8 & 42.1 & 0.015 & 0.089 \\
     & Comet Lake   & 14.2 & 27.7 & 0.050 & 0.190 \\
    \midrule
    \multirow{3}{*}{\rotatebox[origin=c]{90}{\small Sim.}} & ChampSim-pf  & 11.6 & 45.1 & 0.027 & 0.092 \\
     & ChampSim-nopf & 11.7 & 33.2 & 0.053 & 0.178 \\
     & Sniper       & 14.7 & 63.7 & 0.060 & 0.267 \\
    \bottomrule
  \end{tabular}
\end{table}

For every left-out workload, we fit the \ac{bf} regression on the $W{-}1$ remaining workloads, estimate the \ac{bf} at $8\times$ the baseline \ac{mp}, and compute the resulting \ac{cpi}.
We report the error relative to the ground-truth \ac{bf} and the $8\times$~\ac{mp} \ac{cpi} extrapolation.
\autoref{tab:accuracy} shows the $50^{\text{th}}$ and $90^{\text{th}}$ percentile of the relative \ac{cpi} error and the $|\Delta\text{BF}|$.

\begin{figure}[t]
  \centering
  \includegraphics[width=\columnwidth]{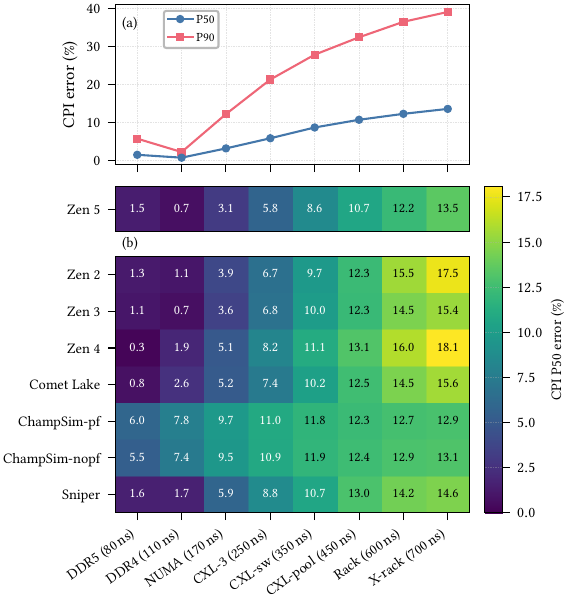}
  \caption{(a)~P50 and P90 \ac{cpi} error on Zen\,5 as a function of the target memory latency. (b)~P50 \ac{cpi} error for each platform (rows) and target latency (columns). Each platform predicts from its own baseline using a single \model{} regression.}
  \label{fig:tech_comparison}
\end{figure}

\begin{figure*}[t!]
  \centering
  \includegraphics[width=\textwidth,height=0.21\textheight,keepaspectratio]{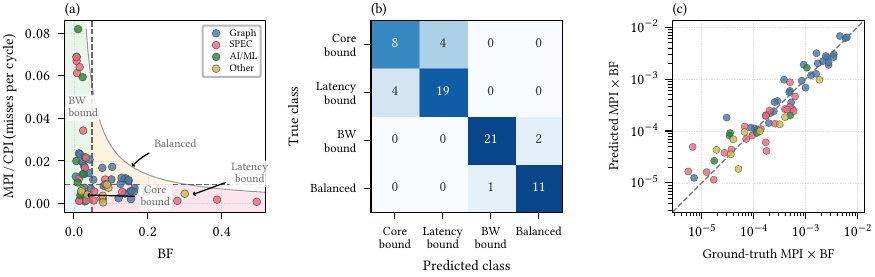}
  \caption{Workload classification and ranking on Zen\,5 (\resLatSensNwl{} workloads). (a)~Classification space. The $x$-axis is the \acs{bf} (latency sensitivity), the $y$-axis is $\mathit{MPI}/\mathit{CPI}$ (demand misses per cycle). Dashed lines mark the median split on each axis. The solid curve is the iso-$f_{\mathrm{mem}}$ hyperbola at the highest observed memory fraction. (b)~Confusion matrix for the four-class partition. Rows are the true class, columns the class predicted by \model{}. (c)~Predicted vs.\ ground-truth $\mathit{MPI} \times \acs{bf}$ (log-log scale).}
  \label{fig:ranking}
\end{figure*}

The \ac{bf} regression is consistent across the four AMD generations: the median $|\Delta\acs{bf}|$ spans \resAmdBfMedLo--\resAmdBfMedHi{} (less than 2\% of the $(0,1)$ \ac{bf} range).
The \ac{cpi} error, however, ranges from \resAmdCpiMedLo\% to \resAmdCpiMedHi\%.
This gap arises because the $\mathit{MPI} \times \mathit{MP}$ product amplifies a small \ac{bf} error into a larger \ac{cpi} error at $8\times$ baseline latency, where the memory fraction $f_{\mathrm{mem}}$ (\autoref{eq:fmem}), the share of the \ac{cpi} attributable to memory, is high and the memory term dominates.
On ChampSim, deterministic counters eliminate run-to-run measurement variation, and the median error drops to \resCsCpiMedLo--\resCsCpiMedHi\% despite higher $|\Delta\acs{bf}|$ (\resCsBfMedLo--\resCsBfMedHi).
The third simulator, Sniper, reaches a median of \resCpiMedSniper\% but the largest 90th percentile in the table.
Sniper also has the largest $|\Delta\acs{bf}|$ at the 90th percentile, and at $8\times$ \ac{mp} that \acs{bf} spread dominates the \ac{cpi} error.
Comet Lake shows the lowest P90 \ac{cpi} error (\resIntelCpiPNinety\%) despite the highest median $|\Delta\acs{bf}|$ (\resIntelBfMed), because its workloads are relatively more core-bound compared to the other platforms.
The 90th-percentile $f_{\mathrm{mem}}$ on Comet Lake is \resFmemPNinetyIntel, compared to \resFmemPNinetyZenFour{} on Zen\,4 and \resFmemPNinetyZenFive{} on Zen\,5.
A lower $f_{\mathrm{mem}}$ reduces the weight of the memory term in the \ac{cpi}, so a \ac{bf} error propagates less into the final prediction.

\autoref{fig:tech_comparison} shows how the \ac{cpi} error changes as the target memory latency increases from DDR5 (80\,ns) to X-rack (700\,ns).
In panel~(a), the P50 error on Zen\,5 rises from \resTechPfiftyFirst\% to \resTechPfiftyLast\%, but the growth saturates, with an increment of about 1\,pp from Rack to X-rack.
As the target \ac{mp} grows, both the predicted and the true \ac{cpi} become dominated by the $\acs{bf} \times \mathit{MPI} \times \mathit{MP}$ term.
The $\mathit{MPI} \times \mathit{MP}$ factor appears in both the numerator and the denominator of the relative error and cancels, so the relative \ac{cpi} error converges toward the relative \ac{bf} error, which is set by the \ac{loo} regression and does not depend on the target latency.
The P90 error increases from \resTechPninetyFirst\% to \resTechPninetyLast\% and does not saturate within this latency range.

An architect evaluating a technology transition needs predictions at absolute latencies.
We therefore perform an extrapolation exercise using generic unloaded latencies of reference memory technologies (e.g., 80\,ns for DDR5, 250\,ns for \ac{cxl}-3, 700\,ns for rack-scale disaggregated memory).
Panel~(b) plots the P50 error for all eight platforms across the technologies.
Because the measured baseline differs across platforms (\resBaseNsIntel--\resBaseNsZenFive\,ns on hardware), a given technology implies a different extrapolation distance on each platform.
For example, DDR5 at 80\,ns falls below the Zen\,5 baseline of 101\,ns, so the prediction there is an interpolation rather than an extrapolation, while DDR4 at 110\,ns sits just above it.
This difference in extrapolation distance explains the non-monotonic errors in the first two columns.
ChampSim starts at higher error because DDR5 already represents ${\approx}2\times$ its baseline, and its error grows slowly, consistent with being closer to its \ac{bf}-error bound.
At CXL-3 latency (${\approx}250$\,ns), the P50 error stays below 12\% on every platform, with Zen\,5 at \resCpiCxldirZenFive\%.

\subsection{Qualitative Workload Analysis}
\label{sec:exp:decisions}

Many architectural decisions depend on knowing which workloads are most sensitive to memory latency rather than on the exact \ac{cpi} of each one.
Typical examples include placing workloads across a tiered memory hierarchy, prioritizing optimization targets, and sizing a fast-memory pool.
We evaluate two complementary measures of qualitative workload behavior.
First, we classify workloads into broad memory-behavior categories, a coarse but actionable partition for capacity planning.
Second, we assess the full pairwise ranking of workloads by predicted sensitivity, a strictly finer-grained measure because any ranking error near a class boundary can flip a classification, but most ranking errors do not.

We place each workload in a two-dimensional space defined by two quantities that can be computed from a single baseline measurement (\autoref{fig:ranking}a).
The $x$-axis is the \ac{bf}, which captures how sensitive a workload is to added memory latency.
The $y$-axis is the ratio $\mathit{MPI}/\mathit{CPI}$, expressed in demand misses per cycle, which measures how frequently the core issues memory requests regardless of how long each request takes.
Together, the two axes separate workloads that generate many misses but tolerate them well (high $\mathit{MPI}/\mathit{CPI}$, low \acs{bf}) from those that generate few misses but stall heavily on each one (low $\mathit{MPI}/\mathit{CPI}$, high \acs{bf}).

We split each axis at its median across the \resLatSensNwl{} Zen\,5 workloads (dashed lines in \autoref{fig:ranking}a), creating four quadrants, each representing a workload class.
\emph{Core-bound} workloads (low on both axes) are the least memory-sensitive of the retained corpus.
\emph{Latency-bound} workloads (high \acs{bf}, low miss rate) stall on misses but issue few of them.
\emph{Band\-width-bound} workloads (low \acs{bf}, high miss rate) generate many misses but overlap them with computation.
\emph{Balanced} workloads (high on both axes) are relatively miss-intensive and stall-prone.
The solid curve marks the iso-$f_{\mathrm{mem}}$ hyperbola at the highest observed memory fraction ($f_{\mathrm{mem}} = \acs{bf} \times \mathit{MPI} \times \mathrm{MP} / \mathit{CPI}$); the region beyond it is unreachable at this operating point.

\autoref{fig:ranking}b shows the confusion matrix for the four classes.
Since $\mathit{MPI}$ and $\mathit{CPI}$ are measured, only the inferred \ac{bf} can move a workload across a class boundary.
\model{} assigns \resConfCorrect{} of the \resLatSensNwl{} workloads (\resConfAccuracy\%) to the correct class, with the \resConfErrors{} misclassifications concentrated near the \ac{bf} median split (core-bound $\leftrightarrow$ latency-bound and BW-bound $\leftrightarrow$ balanced).

The full pairwise ranking is a stricter test: two workloads in the same quadrant can still be misordered without affecting their class label.
We compare the workload ranking produced by \model{} against the ground-truth ranking from \textcite{Clapp:Quantifying:2015}, measured through a full frequency sweep.
We compute Kendall's $\tau$ (\autoref{sec:methodology:eval}) for two predicted quantities at $3\times$~\ac{mp}: the \ac{cpi} and the $\mathit{MPI} \times \acs{bf}$ product.
\autoref{fig:ranking}c shows that the predicted $\mathit{MPI} \times \acs{bf}$ product tracks the ground-truth value along the diagonal across three orders of magnitude, with no systematic bias for either low- or high-sen\-si\-ti\-vi\-ty workloads.
The \ac{cpi}-based ranking achieves $\tau = \resLatSensTauCpiMin$--$\resLatSensTauCpiMax$ across all eight platforms, so fewer than 7\% of pairwise orderings disagree with the ground truth.
Ranking by $\mathit{MPI} \times \acs{bf}$ alone yields a lower $\tau$ (\resLatSensTauMin--\resLatSensTauMax) because this product depends entirely on the inferred \ac{bf}.
The \ac{cpi} ranking is more robust because the predicted \ac{cpi} at a new \ac{mp} adds only the \ac{bf}-dependent delta to the measured baseline \ac{cpi}, and the baseline dominates.
At $8\times$~\ac{mp}, $\tau$ falls to $\resLatSensTauCpiHiMin$--$\resLatSensTauCpiHiMax$, consistent with the larger \ac{cpi} errors at this distance (\autoref{tab:accuracy}).
The ranking degrades but the overall trend is preserved on every platform.


\subsection{When Is HBM Worth It?}
\label{sec:exp:arm}

Consider an architect deciding on the memory configuration of a new platform.
The question is whether adding \mbox{\acs{hbm}} to an existing system provides benefits for a given set of workloads.
As the proposed platform does not yet exist, it is not possible to directly measure and compare the performance benefits.
The common approach to tackle this problem is through simulation of a specific memory configuration.
However, exploring the multiple possible memory configurations (e.g., channels, banks, latency, etc.) and workloads requires a prohibitive amount of resources.
As long as the configuration can be represented through a set of latency-bandwidth response curves, \mbox{\model{}} predicts the performance of a workload on it from a single run on the current system.

To illustrate this scenario, we use a real platform with both \mbox{\acs{hbm}} and \mbox{\acs{ddr}} (see \mbox{\autoref{tab:platforms}}).
The system provides \mbox{\acs{hbm}} in multiple NUMA domains (near and far).

We fit \mbox{\model{}} on the local \mbox{\acs{ddr}} and predict the performance of each workload on near and far \mbox{\acs{hbm}} from that single run.
Reaching the far memory raises the penalty to \mbox{$\sim$\resMeasMpRangeRArmFar$\times$} the baseline.
The far-node memory latency itself falls in the \mbox{\acs{cxl}} range of the \mbox{\acs{ddr}}-to-rack spectrum the model extrapolates over (\mbox{\autoref{fig:tech_comparison}}), so real hardware anchors that extrapolation at a \mbox{\acs{cxl}}-like latency.


We run 16 copies of each benchmark (similar to SPEC rate) so the shared memory system is under realistic bandwidth pressure.

From one \mbox{\acs{ddr}} run, \emph{without ever running the workloads on the target configurations}, \mbox{\model{}} predicts the \mbox{\acs{cpi}} on the near and far \mbox{\acs{hbm}}.
Since we \emph{do} have access to the actual \mbox{\acs{hbm}} memory configurations, we are able to determine the \mbox{\acs{cpi}} prediction accuracy.
\mbox{\model{}} predictions achieve \mbox{\resMeasAnetoPArmFar\%} median and \mbox{\resMeasAnetoPNinetyArmFar\%} 90th percentile error over all the 70 workloads under study.

\begin{figure}[t]
  \centering
  \includegraphics[width=\columnwidth]{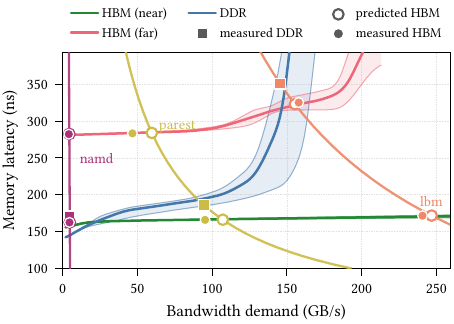}
  \caption{Realistic scenario where \model{} is able to predict the performance of a workload on \mbox{\acs{hbm}} from one run at DDR.
  The shaded bands represent the envelope of memory supply curves from 50:50 to 100:0 read:write ratios.}
  \label{fig:arm_worked}
\end{figure}

\mbox{\autoref{fig:arm_worked}} illustrates the behavior of three representative workloads.
It depicts the predicted and the measured operating points on both \mbox{\acs{hbm}} configurations.

As expected, the bandwidth-bound workloads, represented in the figure by \mbox{\texttt{lbm}}, benefit most from the near \mbox{\acs{hbm}}.
\mbox{\texttt{parest}} only gets a small performance gain (+7\%) when moving to the near \mbox{\acs{hbm}}, while having a significant performance degradation (-47\%) on far \mbox{\acs{hbm}} due to the increase in latency.
On the tested platform, \mbox{\texttt{namd}} is insensitive to memory and thus not greatly affected by memory placement.
Overall, \mbox{\resArmBenefitPctNear\%} of the workloads gain more than 10\% from the near \mbox{\acs{hbm}}.


\subsection{Cross-Platform Analysis}
\label{sec:exp:insight}

We have so far fitted each platform independently.
We now ask whether a single set of regression slopes can serve multiple platforms, with only the intercept~$\beta_0$ absorbing the mean \ac{bf} difference between platforms.

\begin{figure}[t]
  \centering
  \includegraphics[width=\columnwidth]{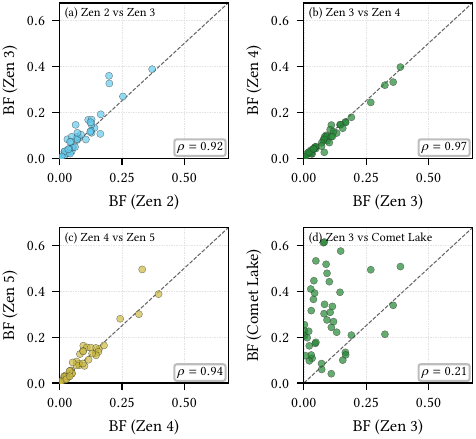}
  \caption{Per-workload \ac{bf} across platform pairs (\resBfNshared{} shared workloads).
  Panels~(a)--(c): adjacent AMD Zen generations; Spearman's~$\rho$ is annotated.
  Panel~(d): Zen\,3 vs.\ Comet Lake.}
  \label{fig:bf_crossgen}
\end{figure}

\autoref{fig:bf_crossgen} plots the per-workload \ac{bf} on one platform against another for the \resBfNshared{} workloads common to both platforms, the intersection of their qualifying sets, since a pairwise comparison needs the same workload on each axis.
Within the AMD Zen family (panels~a--c), the points cluster along the diagonal.

Spearman's $\rho$ exceeds \resBfRhoCrossZenTwoZenThree{} for all three adjacent generation pairs, so a workload that is memory-sensitive on Zen\,2 remains memory-sensitive on Zen\,5 even though the absolute \ac{bf} values may shift with each generation.
Panel~(d) compares Zen\,3 against Comet Lake: $\rho$ drops to \resBfRhoCrossZenThreeIntel{} and the scatter spreads well off the diagonal, so the per-workload \ac{bf} ranking does not transfer across vendors.

An interesting observation is how close the estimated \ac{bf} parameters are across all the Zen generations, while there is a significant difference between similarly aged AMD and Intel machines.
Possible reasons for this range from differences in \acs{pmu} counters to fundamentally different prefetchers or microarchitecture.
This similarity between Zen platforms suggests that a single set of \acs{bf} parameters could model an entire series of machines.
We quantify the prediction impact by pooling the workload data collected from the four AMD generations into a single regression model.
The pooled model shares two beta parameters ($\beta_1^* = \resBetaXshared$, $\beta_2^* = \resBetaVshared$) across all generations and varies only the intercept~$\beta_0$ per platform.
We then apply those AMD-derived slopes to Comet Lake and ChampSim configurations by re-estimating only their intercepts.

We compare the accuracy of the per-platform latency sensitivity to the pooled model (with only per-platform $\beta_0$ values).
On all four AMD generations, the shared-slope error matches the per-platform fit to within $1{\times}10^{-4}$ at the median and $7{\times}10^{-4}$ at the 90th percentile.
The Zen\,5 P90, for example, moves from $4.8{\times}10^{-3}$ to $4.9{\times}10^{-3}$, a negligible change.
Outside the Zen family, the gap grows with architectural distance.
The Comet Lake P90 rises from $2.2{\times}10^{-3}$ to $3.0{\times}10^{-3}$ (a 36\% increase).

Within a microarchitectural family the \ac{bf} ranking seems to be preserved and shared slopes may suffice.
Across vendors or fundamentally different memory subsystems, per-platform coefficients are necessary.

\subsection{Accuracy Limits and Model Cost}
\label{sec:exp:limits}

This section covers two important remaining questions. How does prediction accuracy vary across different workloads? How much measurement effort does \model{} require?

\begin{figure}[t]
  \centering
  \includegraphics[width=\columnwidth]{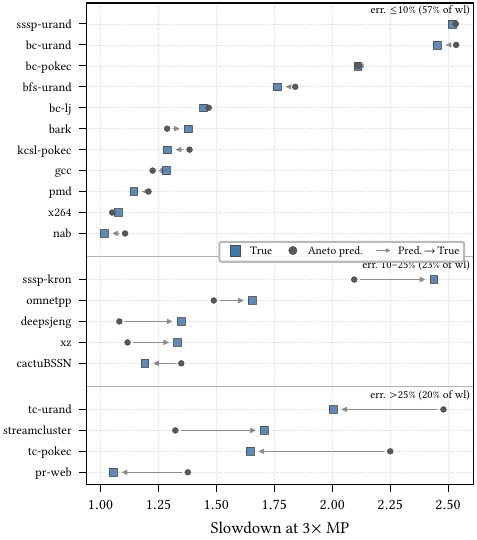}
  \caption{Ground truth (squares) vs.\ \model{} predicted (circles) slowdown under a $3\times$ \ac{mp} extrapolation on Zen\,5.}
  \label{fig:top_workloads}
\end{figure}

We address the first question in \autoref{fig:top_workloads}.
We predict the slowdown at $3\times$~\ac{mp} on Zen\,5 for all \resTopNwl{} workloads and compare against the ground truth.
The figure shows \resTopNshow{} representative workloads sampled across the error distribution.
Of the \resTopNwl{} workloads, \resTopPctGood\% fall within 10\% of the ground truth, \resTopPctModerate\% show a moderate error (10--25\%), and \resTopPctChallenging\% exceed 25\%.
The \resTopPctChallenging\% challenging cases share two traits: a high memory fraction $f_{\mathrm{mem}}$ and a relatively low per-workload $R^2$ in the linear \ac{cpi} decomposition, still above the \mbox{$0.90$} retention gate.
A low $R^2$ means that the \ac{cpi} of these workloads does not follow the linear $\mathit{CPI}_0 + \acs{bf} \times \mathit{MPI} \times \mathit{MP}$ relationship even when fitted individually.
The constant-\ac{bf} assumption is not sufficient to fully describe their memory behavior.
In these cases there are several mechanisms (e.g., the memory access patterns, execution phases, or locks) that may challenge the assumptions.
The high $f_{\mathrm{mem}}$ then amplifies the resulting \ac{bf} error into a large \ac{cpi} error, as described in \autoref{sec:exp:accuracy}.
For the \resTopPctGood\% majority, the prediction is accurate enough to shortlist workloads for detailed cycle-level simulation.

\subsubsection{Minimal Workload Set}
\label{sec:exp:corpus}

Using the \model{} regression model on a new platform requires a minimal number of diverse workloads to fit the $\beta$ parameters.
In this section we determine how many workloads are needed by subsampling.

On each of the \resBetaStabNplatforms{} platforms, we draw \resBetaStabM{} random subsets of increasing size ($w = 3, 4, \ldots, W$) from the available workloads.
For each subset, we fit the regression and measure the \ac{bf} prediction error on the remaining workloads.
We define $W^*$ as the smallest $w$ at which the median held-out error is within \resBetaStabDelta{} of the error obtained when using all workloads (the \ac{loo} error).
\autoref{tab:beta_stab} reports the result. Here $W$ is the number of workloads common to all hardware platforms or to all simulators, and $W^*$ ranges from \resBetaStabNstarMin{} to \resBetaStabNstarMax{} across the eight platforms, with a median of \resBetaStabNstarMedian.
Because the regression has only three parameters, a small corpus suffices, though the worst case (\resBetaStabNstarMax{} on ChampSim-nopf) is larger than the median.

\begin{table}[b]
  \centering
  \caption{Corpus size ($W^*$) and \ac{cpi} P50 error at $3\times$\,\ac{mp}: 16 fixed workloads vs.\ full \ac{loo}.}
  \label{tab:beta_stab}
  \small
  \begin{tabular}{lrrcccr}
    \toprule
    & & & \multicolumn{3}{c}{CPI P50 (3$\times$ \ac{mp})} \\
    \cmidrule(lr){4-6}
    Platform & $W$ & $W^*$ & 16 & LOO & $\Delta$ \\
    \midrule
    Zen\,2 & 44 & 5 & 8.1\% & 9.4\% & -1.3 \\
    Zen\,3 & 44 & 5 & 8.1\% & 5.9\% & +2.1 \\
    Zen\,4 & 44 & 5 & 9.1\% & 9.6\% & -0.5 \\
    Zen\,5 & 44 & 5 & 6.2\% & 7.3\% & -1.1 \\
    Comet Lake & 44 & 6 & 4.2\% & 6.6\% & -2.4 \\
    ChampSim-pf & 84 & 6 & 7.5\% & 8.3\% & -0.9 \\
    ChampSim-nopf & 84 & 20 & 8.0\% & 7.7\% & +0.3 \\
    Sniper & 84 & 7 & 12.4\% & 10.0\% & +2.3 \\
    \bottomrule
  \end{tabular}
\end{table}

We round up to \resBetaStabK{} workloads, selected once on a reference platform to cover low, mid, and high \ac{bf} values, and apply the same set unchanged to every target platform.
\autoref{tab:beta_stab} compares the \ac{cpi} P50 error at $\resBetaStabMpFactor\times$~\ac{mp} (a DDR-to-\ac{cxl} extrapolation) between this fixed corpus and the full \ac{loo} protocol.
The difference ($\Delta$) stays within $\pm\resBetaStabDeltaAbsMax$\,pp on all eight platforms, so fixing the corpus at \mbox{\resBetaStabK{}} workloads tracks the full \mbox{\acs{loo}} error closely and never incurs a systematic penalty.
Each target platform runs the frequency sweep on the same \resBetaStabK{} workloads to obtain its own regression coefficients, but the workload selection is done once.

\subsubsection{Modeling Cost}

With the minimal workload set fixed at \resBetaStabK{} workloads, we now quantify the total measurement effort and compare it against Clapp~\etal{}~\cite{Clapp:Simple:2015}.

Clapp~\etal{}~\cite{Clapp:Simple:2015} require a full \ac{dvfs} sweep for every workload to fit the per-workload linear model, costing a median of \resClappMarginalZenFive{}~minutes per workload on Zen\,5.
The total measurement time therefore grows linearly with the number of workloads characterized.
\model{} instead pays a one-time cost: the \resCostTrainingWl{} corpus workloads run the full frequency sweep (\resCalibrationCostZenFive{}~machine-hours on Zen\,5).
After this, each additional workload requires only a single measurement at the baseline frequency (\resAnetoCostZenFive{}~minutes), a \resCostReductionZenFive{}$\times$ reduction in per-workload cost.

In sum, \model{} trades a bounded calibration investment for unbounded marginal savings: once \resBetaStabK{} workloads have been swept, every subsequent characterization reduces to a single instrumented workload and a sub-second regression query.

\subsection{Comparison with PROFET}
\label{sec:exp:profet}

\begin{table}[b]
  \centering
  \caption{\model{} vs.\ PROFET: \ac{cpi} error (\%) on ChampSim.}
  \label{tab:profet}
  \small
  \begin{tabular*}{\columnwidth}{@{\extracolsep{\fill}}llcccc}
    \toprule
    & & \multicolumn{2}{c}{PROFET} & \multicolumn{2}{c}{\model{}} \\
    \cmidrule(lr){3-4} \cmidrule(lr){5-6}
    Platform & \ac{mp} & P50 & P90 & P50 & P90 \\
    \midrule
    \multirow{3}{*}{ChampSim-pf} & $2\times$ MP & 7.8 & 34.3 & \textbf{4.4} & \textbf{17.7} \\
     & $3\times$ MP & 12.9 & 58.1 & \textbf{6.8} & \textbf{32.1} \\
     & $8\times$ MP & 16.9 & 81.7 & \textbf{10.2} & \textbf{51.0} \\
    \midrule
    \multirow{3}{*}{ChampSim-nopf} & $2\times$ MP & 6.4 & 43.8 & \textbf{5.4} & \textbf{12.2} \\
     & $3\times$ MP & 9.6 & 82.2 & \textbf{8.2} & \textbf{21.3} \\
     & $8\times$ MP & 12.1 & 142.2 & \textbf{11.3} & \textbf{32.7} \\
    \bottomrule
  \end{tabular*}
\end{table}

Both \model{} and PROFET~\cite{Radulovic:PROFET:2019} predict the \ac{cpi} at an unseen memory latency from a single baseline measurement, making PROFET the closest point of comparison\footnote{We use the reference implementation at \url{https://github.com/bsc-mem/PROFET}, commit \texttt{ae0157a}.}.
We evaluate both on the two ChampSim configurations under identical conditions: the same workload corpus, the same ground-truth quality threshold ($R^2 \geq 0.90$, \autoref{sec:methodology:gt}), and constant-latency memory curves (no queuing delay).
These conditions favor PROFET: there is no bandwidth contention, and each ChampSim trace covers 250 million instructions, a far shorter window than the one-second sampling intervals that PROFET uses on real hardware.
The shorter window reduces phase variability within each sample, so the uniform-behavior assumption that underpins the PROFET \ac{mlp} estimate holds more closely.

\autoref{tab:profet} reports the \ac{cpi} prediction error at three extrapolation distances.
\model{} achieves \resPfXrangeCsDpc{}$\times$ lower P50 error than PROFET on ChampSim-pf and \resPfXrangeCsNopf{}$\times$ lower on ChampSim-nopf.
The separation is wider in the tail, where the median P90 error is \resPfPNMedCsDpc{}$\times$ lower on ChampSim-pf and \resPfPNMedCsNopf{}$\times$ lower on ChampSim-nopf.
To confirm that this advantage is not driven by a few outliers, we run a one-sided Wilcoxon signed-rank test on the paired per-workload absolute errors.
The null hypothesis, that \model{} has equal or higher error than PROFET, is rejected at $8\times$~\ac{mp} on both platforms ($p \resPfPEightCsDpc$ on ChampSim-pf, $p \resPfPEightCsNopf$ on ChampSim-nopf).
We conclude that \model{} produces systematically lower error than PROFET on the majority of individual workloads.

This accuracy difference is due to two properties of the PROFET analytical model.
After an \ac{llc} miss, the out-of-order engine continues executing independent instructions ($\mathit{Ins}_{\mathrm{OOO}}$) until the \ac{rob} fills or the data arrives.
PROFET cannot measure $\mathit{Ins}_{\mathrm{OOO}}$ directly, so it treats this quantity as a free parameter bounded by the \ac{rob} size and the miss penalty.
It sweeps $\mathit{Ins}_{\mathrm{OOO}}$ within these bounds to derive the \ac{mlp}, reports the mean prediction over the sweep, and returns the endpoints as a sensitivity band.
That band spans \resPfBandRange\% of the predicted \ac{cpi} and contains the measured value for \resPfCoverRange\% of workloads.
The procedure can bound the \ac{bf} but cannot determine it from the measured counters.

A second limitation is sensitivity to the \ac{llc} miss count, which must capture demand misses only.
On real processors the available counters often include prefetch traffic, and because PROFET derives its prediction from a single per-workload ratio of miss counts to instructions, any bias propagates directly into the predicted \ac{cpi}.
We confirm this on ChampSim-pf, where feeding PROFET demand-plus-prefetch counters raises its P50 error from \resPfDemPfiftyThree\% to \resPfAllPfiftyThree\% at $3\times$~\ac{mp} and from \resPfDemPfiftyEight\% to \resPfAllPfiftyEight\% at $8\times$.

\model{} avoids both limitations.
We estimate the \ac{bf} from the full workload corpus with no per-workload free parameter.
If the \ac{mpi} counters carry a systematic bias (e.g., an \ac{llc}-slice \ac{pmu} that cannot separate demand from prefetch misses), the \ac{ols} coefficients shift during fitting to map the inflated inputs to the correct ground-truth \ac{bf}.
Because the same inflation is present at prediction time, the shift compensates and \model{} maintains comparable accuracy with either \ac{mpi} definition.

\subsection{Model Limitations}
\label{sec:exp:limitations}

\model{} inherits the assumptions and limitations of the Interval Models family on which it builds~\mbox{\cite{Eyerman::Mechanistic:2009}}.
The strongest limitation is that the \mbox{\acs{bf}} remains constant as the \mbox{\acs{mp}} varies.
This can break for workloads whose memory behavior shifts with latency, though we did not observe it to degrade prediction accuracy in practice (\mbox{\autoref{sec:model:scope}}).
The cross-workload regularity also does not explain the full \mbox{\acs{bf}} variation.
This is shown in \mbox{\autoref{fig:bf_gradient}} where the \mbox{\acs{bf}} of some workloads (color shaded) deviated from the diagonal bands.
This may be driven by phase behavior, irregular access patterns, or operating system interactions that \model{} does not capture.
Accuracy rests on the quality of the underlying measurements, especially the miss penalty.
Ultimately, inaccurate latency-bandwidth response curves limit the predictive power.

\section{Related Work}
\label{sec:related}

Analytical performance models span a mechanistic-to-empirical spectrum~\cite{Eeckhout:PerfEvalMethods:2010}, trading modeling assumptions for measurement requirements.
We position \model{} relative to both traditions and to memory-specific models.

\subsection{Mechanistic Models}
\label{sec:related:mechanistic}

The additive \ac{cpi} decomposition, separating on-chip cost from memory stall penalty, originates with Emma~\cite{Emma:Understanding:1997} and was formalized as \ac{cpi} stacks by Eyerman~\etal{}~\cite{Eyerman:CPIComponents:2006}.
Karkhanis and Smith~\cite{Karkhanis:Firstorder:2004} modeled the decomposition for superscalar \ac{ooo} processors, showing that the effective miss penalty depends on how much useful work the pipeline overlaps during the miss~\cite{Karkhanis:DayInTheLife:2002}.
The \ac{bf} reformulates this per-miss overlap fraction as a single scalar that captures the fraction of memory latency the processor fails to hide.

The \textit{Interval Model}~\cite{Eyerman::Mechanistic:2009} generalized this by dividing execution into base and penalty intervals, amortizing the penalty of co-oc\-cur\-ring \ac{llc} misses.
Its sub-linear scaling with miss rate, bounded by \ac{rob} depth and \ac{mshr} count, is the structure the \ac{bf} captures and \model{} exploits.
Because memory latency is an explicit parameter, the Interval Model can predict the performance under a different \ac{mp}, but its required inputs (per-workload miss profiles, \ac{rob} occupancy, \ac{mlp} distributions) must be extracted from costly traces or time-consuming simulation.
Van den Steen~\etal{}~\cite{VanDenSteen:MicroarchitectureIndependent:2016} reduced the profiling cost by collecting a micro-architecture independent profile through dynamic binary instrumentation~\cite{Hoste:MicroarchIndependent:2007}.
From a single profile, the model predicts across cache, \ac{rob}, branch predictor, and memory latency configurations.
Each workload still requires an instrumentation pass.
\model{} is specialized for memory-latency sensitivity prediction and trades generality for measurement simplicity: three \ac{pmu} counters from a single baremetal run, with no workload instrumentation.

\subsection{Empirical Models}
\label{sec:related:empirical}

Empirical design-space models~\cite{Ipek:DSE:2006, Joseph:Regression:2006, Lee:RegressionModeling:2006} learn performance surfaces over \textit{microarchitectural configurations} from training data; each new design point requires retraining.
More recent ML-based approaches fuse analytical templates with learned corrections~\cite{NasrEsfahany:Concorde:2025}, but they require large training corpora that are typically generated through cycle-accurate simulation~\cite{Binkert:gem5:2011}.
The training labels therefore inherit the simulator's own modeling error, which can be 5--15\% even after careful calibration~\cite{Gutierrez:Sources:2014}.
\model{} side\-steps both the data hunger and the error floor: it fits a three-co\-ef\-fi\-cient regression model from baremetal \ac{pmu} readings on real hardware.

\model{} retains the mechanistic structure (\autoref{eq:cpi_intro}) but estimates the \ac{bf} by exploiting a cross-workload regularity that neither these empirical models nor the mechanistic models above exploit.
\subsection{Memory Performance Models}
\label{sec:related:memory}

MESS~\cite{Esmaili:MESS:2024} characterizes a platform's bandwidth-latency supply curves through synthetic benchmarks (\autoref{sec:motivation}).
MESS is orthogonal to \model{} and is synergistic: MESS provides the memory-side characterization while \model{} provides the per-workload \ac{bf} coefficients.
On the demand side, Clapp~\etal{}~\cite{Clapp:Simple:2015, Clapp:Quantifying:2015} reformulated the \ac{bf} decomposition (\autoref{eq:cpi_intro}) by Chou~\etal{}~\cite{Chou:MLP:2004} and showed that \ac{bf} and $\mathit{CPI}_0$ can be extracted by sweeping the effective \ac{mp} across frequency steps.
We use Clapp's multi-point regression model as the ground truth throughout this paper.
PROFET~\cite{Radulovic:PROFET:2019} avoids the sweep by treating the per-miss overlap as a free parameter, but this produces a sensitivity band rather than a point estimate, and it relies on accurate demand \ac{mpi}.
We compare both methods in \autoref{sec:exp:profet}.

\section{Conclusions}
\label{sec:conclusions}

We observe a cross-workload regularity: on a given platform, the latent \ac{bf} can be inferred from the behavior of other workloads on the same machine, given only their \ac{cpi}, \ac{mpi}, and \ac{mp}, without per-workload profiling, sweeps, or simulation.
\model{} demonstrates that this regularity is practically exploitable, predicting \ac{cpi} under arbitrary memory configurations with minimal measurement effort.
Across five physical platforms and three simulator configurations, the median \mbox{\acs{cpi}} error across the eight platforms is \mbox{\resCpiAll{}\%} (\autoref{tab:accuracy}) even at $8\times$ \ac{mp} extrapolation, comparable to the 5--15\% accuracy floor of cycle-accurate simulators~\cite{Gutierrez:Sources:2014} and \resPfPNRatio{}$\times$ lower 90th-percentile error than the current state of the art~\cite{Radulovic:PROFET:2019} on the two ChampSim configurations.
On an ARM server, we compare the predictions against real measurements at up to \mbox{$\sim$\resMeasMpRangeRArmFar$\times$} the baseline memory penalty and under a different memory technology (\mbox{\acs{hbm}}). The median \mbox{\acs{cpi}} error stays at \mbox{\resMeasAnetoPArmFar\%}.
Because \model{} captures the demand side of \ac{cpi} in isolation, it pairs naturally with the supply-side characterization tools such as MESS~\cite{Esmaili:MESS:2024}.
Combining both, an architect can predict how a workload performs on a memory subsystem where it was never executed, using only a single run on the current system.

Several future directions remain open for further investigation.
A deeper analysis of workload characteristics and the microarchitectural implications can uncover the mechanisms that drive the cross-workload regularity identified in this work.
Understanding how the \mbox{\acs{bf}} varies between workloads, CPU generations, and vendors may enable first-order performance prediction on new architectures without running the workloads on the target system.
Extending the same first-order modeling to complex multi-threaded applications would capture the behavior of thread synchronization or I/O.
Composing \mbox{\model{}} with a memory-power model would expand it from performance-only forecasting to joint performance and energy prediction.


\balance
\printbibliography

@article{Besta:GMS:2021,
  author       = {Maciej Besta and Zur Vonarburg-Shmaria and Yannick Schaffner and Leonardo Schwarz and Grzegorz Kwasniewski and Lukas Gianinazzi and Jakub Ber{\'{a}}nek and Kacper Janda and Tobias Holenstein and Sebastian Leisinger and Peter Tatkowski and Esref {\"{O}}zdemir and Adrian Balla and Marcin Copik and Philipp Lindenberger and Marek Konieczny and Onur Mutlu and Torsten Hoefler},
  title        = {{GraphMineSuite}: Enabling High-Performance and Programmable Graph Mining Algorithms with Set Algebra},
  journal      = {Proc. {VLDB} Endow.},
  volume       = {14},
  number       = {11},
  pages        = {1922--1936},
  year         = {2021},
  doi          = {10.14778/3476249.3476252},
}

@INPROCEEDINGS{Akram:Validation:2019,
	author={Akram, Ayaz and Sawalha, Lina},
	booktitle={2019 IEEE/ACM Performance Modeling, Benchmarking and Simulation of High Performance Computer Systems (PMBS)},
	title={Validation of the gem5 Simulator for x86 Architectures},
	year={2019},
	volume={},
	number={},
	pages={53-58},
	doi={10.1109/PMBS49563.2019.00012}
}

@article{Binkert:gem5:2011,
	author = {Binkert, Nathan and Beckmann, Bradford and Black, Gabriel and Reinhardt, Steven K. and Saidi, Ali and Basu, Arkaprava and Hestness, Joel and Hower, Derek R. and Krishna, Tushar and Sardashti, Somayeh and Sen, Rathijit and Sewell, Korey and Shoaib, Muhammad and Vaish, Nilay and Hill, Mark D. and Wood, David A.},
	title = {The gem5 simulator},
	year = {2011},
	issue_date = {May 2011},
	publisher = {Association for Computing Machinery},
	address = {New York, NY, USA},
	volume = {39},
	number = {2},
	issn = {0163-5964},
	url = {https://doi.org/10.1145/2024716.2024718},
	doi = {10.1145/2024716.2024718},
	journal = {SIGARCH Comput. Archit. News},
	month = aug,
	pages = {1–7},
	numpages = {7}
}

@inproceedings{Clapp:Simple:2015,
	author = {Clapp, Russell and Dimitrov, Martin and Kumar, Karthik and Viswanathan, Vish and Willhalm, Thomas},
	title = {A Simple Model to Quantify the Impact of Memory Latency and Bandwidth on Performance},
	year = {2015},
	isbn = {9781450334860},
	publisher = {Association for Computing Machinery},
	address = {New York, NY, USA},
	url = {https://doi.org/10.1145/2745844.2745900},
	doi = {10.1145/2745844.2745900},
	booktitle = {Proceedings of the 2015 ACM SIGMETRICS International Conference on Measurement and Modeling of Computer Systems},
	pages = {471–472},
	numpages = {2},
	location = {Portland, Oregon, USA},
	series = {SIGMETRICS '15},
}

@INPROCEEDINGS{Clapp:Quantifying:2015,
	author={Clapp, Russell and Dimitrov, Martin and Kumar, Karthik and Viswanathan, Vish and Willhalm, Thomas},
	booktitle={2015 IEEE International Symposium on Workload Characterization},
	title={Quantifying the Performance Impact of Memory Latency and Bandwidth for Big Data Workloads},
	year={2015},
	volume={},
	number={},
	pages={213-224},
	doi={10.1109/IISWC.2015.32},
}

@inproceedings{Carlson:Sniper:2011,
	author={Carlson, Trevor E. and Heirman, Wim and Eeckhout, Lieven},
	booktitle={Proceedings of 2011 International Conference for High Performance Computing, Networking, Storage and Analysis (SC '11)},
	title={Sniper: Exploring the Level of Abstraction for Scalable and Accurate Parallel Multi-Core Simulation},
	year={2011},
	pages={52:1--52:12},
	publisher={ACM},
	doi={10.1145/2063384.2063454},
}

@inproceedings{Esmaili:MESS:2024,
	author = {Esmaili-Dokht, Pouya and Sgherzi, Francesco and Girelli, Val\'{e}ria Soldera and Boixaderas, Isaac and Carmin, Mariana and Monemi, Alireza and Armejach, Adri\`{a} and Mercadal, Estanislao and Llort, Germ\'{a}n and Radojkovi\'{c}, Petar and Moreto, Miquel and Gim\'{e}nez, Judit and Martorell, Xavier and Ayguad\'{e}, Eduard and Labarta, Jesus and Confalonieri, Emanuele and Dubey, Rishabh and Adlard, Jason},
	title = {A Mess of Memory System Benchmarking, Simulation and Application Profiling},
	year = {2024},
	publisher = {IEEE Press},
	url = {https://doi.org/10.1109/MICRO61859.2024.00020},
	doi = {10.1109/MICRO61859.2024.00020},
	booktitle = {Proceedings of the 2024 57th IEEE/ACM International Symposium on Microarchitecture},
	pages = {136–152},
	numpages = {17},
	location = {Austin, TX, USA},
	series = {MICRO '24}
}

@article{Eyerman::Mechanistic:2009,
	author = {Eyerman, Stijn and Eeckhout, Lieven and Karkhanis, Tejas and Smith, James E.},
	title = {A mechanistic performance model for superscalar out-of-order processors},
	year = {2009},
	issue_date = {May 2009},
	publisher = {Association for Computing Machinery},
	address = {New York, NY, USA},
	volume = {27},
	number = {2},
	issn = {0734-2071},
	url = {https://doi.org/10.1145/1534909.1534910},
	doi = {10.1145/1534909.1534910},
	journal = {ACM Trans. Comput. Syst.},
	month = may,
	articleno = {3},
	numpages = {37}
}

@INPROCEEDINGS{Gutierrez:Sources:2014,
	author={Gutierrez, Anthony and Pusdesris, Joseph and Dreslinski, Ronald G. and Mudge, Trevor and Sudanthi, Chander and Emmons, Christopher D. and Hayenga, Mitchell and Paver, Nigel},
	booktitle={2014 IEEE International Symposium on Performance Analysis of Systems and Software (ISPASS)},
	title={Sources of error in full-system simulation},
	year={2014},
	volume={},
	number={},
	pages={13-22},
	doi={10.1109/ISPASS.2014.6844457}}

@INPROCEEDINGS{Karkhanis:Firstorder:2004,
  author={Karkhanis, T.S. and Smith, J.E.},
  booktitle={Proceedings. 31st Annual International Symposium on Computer Architecture, 2004.},
  title={A first-order superscalar processor model},
  year={2004},
  volume={},
  number={},
  pages={338-349},
  doi={10.1109/ISCA.2004.1310786}
}

@article{Radulovic:PROFET:2019,
	author = {Radulovic, Milan and S\'{a}nchez Verdejo, Rommel and Carpenter, Paul and Radojkovi\'{c}, Petar and Jacob, Bruce and Ayguad\'{e}, Eduard},
	title = {PROFET: Modeling System Performance and Energy Without Simulating the CPU},
	year = {2019},
	issue_date = {June 2019},
	publisher = {Association for Computing Machinery},
	address = {New York, NY, USA},
	volume = {3},
	number = {2},
	url = {https://doi.org/10.1145/3341617.3326149},
	doi = {10.1145/3341617.3326149},
	journal = {Proc. ACM Meas. Anal. Comput. Syst.},
	month = jun,
	articleno = {34},
	numpages = {33}
}

@ARTICLE{VanDenSteen:MicroarchitectureIndependent:2016,
  author={Van den Steen, Sam and Eyerman, Stijn and De Pestel, Sander and Mechri, Moncef and Carlson, Trevor E. and Black-Schaffer, David and Hagersten, Erik and Eeckhout, Lieven},
  journal={IEEE Transactions on Computers},
  title={Analytical Processor Performance and Power Modeling Using Micro-Architecture Independent Characteristics},
  year={2016},
  volume={65},
  number={12},
  pages={3537-3551},
  doi={10.1109/TC.2016.2547387}
}

@INPROCEEDINGS{Bienia:PARSEC:2008,
	author = {Bienia, Christian and Kumar, Sanjeev and Singh, Jaswinder Pal and Li, Kai},
	title = {The {PARSEC} Benchmark Suite: Characterization and Architectural Implications},
	booktitle = {Proc. PACT},
	year = {2008},
	pages = {72--81}
}

@INPROCEEDINGS{Beamer:GAP:2015,
	author = {Beamer, Scott and Asanovi{\'c}, Krste and Patterson, David},
	title = {The {GAP} Benchmark Suite},
	year = {2015},
	archiveprefix = {arXiv},
	eprint = {1508.03619}
}

@INPROCEEDINGS{Shun:Ligra:2013,
	author = {Shun, Julian and Blelloch, Guy E.},
	title = {Ligra: A Lightweight Graph Processing Framework for Shared Memory},
	booktitle = {Proc. PPoPP},
	year = {2013},
	pages = {135--146}
}

@INPROCEEDINGS{Blackburn:DaCapo:2006,
	author = {Blackburn, Stephen M. and Garner, Robin and Hoffmann, Chris and Khang, Asjad M. and McKinley, Kathryn S. and Bentzur, Rotem and Diwan, Amer and Feinberg, Daniel and Frampton, Daniel and Guyer, Samuel Z. and Hirzel, Martin and Hosking, Antony and Jump, Maria and Lee, Han and Moss, J. Eliot B. and Phansalkar, Aashish and Stefanovi{\'c}, Darko and VanDrunen, Thomas and von Dincklage, Daniel and Wiedermann, Ben},
	title = {The {DaCapo} Benchmarks: {Java} Benchmarking Development and Analysis},
	booktitle = {Proc. OOPSLA},
	year = {2006},
	pages = {169--190}
}

@INPROCEEDINGS{Gober:ChampSim:2022,
	author = {Gober, Nathan and Chacon, Gino and Wang, Lei and Gratz, Paul V. and Jim{\'e}nez, Daniel A. and Teran, Elvira and Pugsley, Seth and Kim, Jinchun},
	title = {The Championship Simulator: Architectural Simulation for Education and Competition},
	year = {2022},
	archiveprefix = {arXiv},
	eprint = {2210.14324}
}

@BOOK{Eeckhout:PerfEvalMethods:2010,
	author = {Eeckhout, Lieven},
	title = {Computer Architecture Performance Evaluation Methods},
	publisher = {Morgan \& Claypool},
	year = {2010},
	series = {Synthesis Lectures on Computer Architecture}
}

@INPROCEEDINGS{Joseph:Regression:2006,
	author = {Joseph, P. J. and Vaswani, Kapil and Thazhuthaveetil, Matthew J.},
	title = {Construction and Use of Linear Regression Models for Processor Performance Analysis},
	booktitle = {Proc. HPCA},
	year = {2006},
	pages = {99--108}
}

@INPROCEEDINGS{Lee:RegressionModeling:2006,
	author = {Lee, Benjamin C. and Brooks, David M.},
	title = {Accurate and Efficient Regression Modeling for Microarchitectural Performance and Power Prediction},
	booktitle = {Proc. ASPLOS},
	year = {2006},
	pages = {185--194}
}

@INPROCEEDINGS{Ipek:DSE:2006,
	author = {Ipek, Engin and McKee, Sally A. and Caruana, Rich and de Supinski, Bronis R. and Schulz, Martin},
	title = {Efficiently Exploring Architectural Design Spaces via Predictive Modeling},
	booktitle = {Proc. ASPLOS},
	year = {2006},
	pages = {195--206}
}

@inproceedings{NasrEsfahany:Concorde:2025,
	author = {Nasr-Esfahany, Arash and Alizadeh, Mohammad and Lee, Victor and Alam, Hanna and Coon, Brett W. and Culler, David E. and Dadu, Vidushi and Dixon, Martin and Levy, Henry M. and Pandey, Santosh and Ranganathan, Parthasarathy and Yazdanbakhsh, Amir},
	title = {Concorde: Fast and Accurate {CPU} Performance Modeling with Compositional Analytical-{ML} Fusion},
	booktitle = {Proc. ISCA},
	year = {2025},
	pages = {1480--1494},
	doi = {10.1145/3695053.3731037}
}

@ARTICLE{Emma:Understanding:1997,
	author = {Emma, Philip G.},
	title = {Understanding Some Simple Processor-Performance Limits},
	journal = {IBM Journal of Research and Development},
	volume = {41},
	number = {3},
	pages = {215--232},
	year = {1997}
}

@INPROCEEDINGS{Eyerman:CPIComponents:2006,
	author = {Eyerman, Stijn and Eeckhout, Lieven and Karkhanis, Tejas and Smith, James E.},
	title = {A Performance Counter Architecture for Computing Accurate {CPI} Components},
	booktitle = {Proc. ASPLOS},
	year = {2006},
	pages = {175--184}
}

@INPROCEEDINGS{Bucek:SPEC2017:2018,
	author = {Bucek, James and Lange, Klaus-Dieter and von Kistowski, J{\'o}akim},
	title = {{SPEC CPU2017}: Next-Generation Compute Benchmark},
	booktitle = {Proc. ICPE Companion},
	year = {2018},
	pages = {41--42}
}

@INPROCEEDINGS{Karkhanis:DayInTheLife:2002,
	author = {Karkhanis, Tejas S. and Smith, James E.},
	title = {A Day in the Life of a Data Cache Miss},
	booktitle = {Proc. Workshop on Memory Performance Issues (WMPI)},
	year = {2002}
}

@INPROCEEDINGS{Hoste:MicroarchIndependent:2007,
	author = {Hoste, Kenneth and Eeckhout, Lieven},
	title = {Microarchitecture-Independent Workload Characterization},
	booktitle = {IEEE Micro},
	volume = {27},
	number = {3},
	pages = {63--72},
	year = {2007}
}

@inproceedings{Chou:MLP:2004,
  author       = {Chou, Yuan and Fahs, Brian and Abraham, Santosh G.},
  title        = {Microarchitecture Optimizations for Exploiting Memory-Level Parallelism},
  booktitle    = {31st International Symposium on Computer Architecture (ISCA)},
  pages        = {76--89},
  publisher    = {{IEEE}},
  year         = {2004},
  doi          = {10.1109/ISCA.2004.1310765}
}

\clearpage
\raggedbottom 
\renewcommand{\appendixname}{Artifact Appendix}
\appendix

\subsection{Abstract}

This artifact reproduces every results figure, table, and number quoted in
the text of the paper, except the ARM results (see Notes). The artifact contains the data used to fit and evaluate the model, from
hardware measurements and simulation, and the code that analyzes it. The data
are the \emph{aneto-metrics-v1} JSON files for all eight platforms (about
4\,MB). Each file holds, for one benchmark suite on one
platform, the hardware performance-counter measurements, or the simulator
statistics on ChampSim and Sniper, that the analysis needs. The code is a
reference implementation of the \model{} model in about 1.5k lines of Python.
Each figure, table, or set of quoted numbers has
its own script that reproduces it from the data. A verifier then compares
every reproduced value against the reference outputs shipped in the
\texttt{expected/} directory. Because the analysis starts from the included
measurements, no special hardware is required. The measurements themselves are
not re-collected, which would need access to the eight platforms and several
days of benchmark builds and measurement sweeps. Reproducing and verifying
everything from the included measurements requires less than five minutes on
a commodity x86-64 machine.

\subsection{Artifact check-list (meta-information)}

{\small
\begin{itemize}
  \item {\bf Algorithm:} Clapp~\etal{} multi-point regression, \model{}
        \acs{bf} regression, \acs{loo} cross-validation.
  \item {\bf Program:} Python 3.12, no compiled code.
  \item {\bf Data set:} aneto-metrics-v1 JSON metrics for 8 platforms
        (5 hardware \acs{dvfs} sweeps, 3 simulator latency sweeps), included.
  \item {\bf Run-time environment:} Linux, Python 3.12 virtualenv with
        pinned packages, TeX~Live for figure fonts, Dockerfile included.
  \item {\bf Hardware:} any x86-64 machine with 8\,GB of RAM or more.
  \item {\bf Execution:} \texttt{scripts/run\_all.sh}, less than five minutes.
  \item {\bf Metrics:} relative \acs{cpi} error percentiles (P50 and P90)
        at each \acs{mp} factor, absolute \acs{bf} error, $R^2$, Kendall's $\tau$,
        Spearman's $\rho$.
  \item {\bf Output:} \texttt{experiments/*/out/} macros, tables, and
        figure PDFs, checked by \texttt{scripts/verify.py}.
  \item {\bf Experiments:} 11 Makefile-driven scripts, one per figure,
        table, or set of quoted numbers.
  \item {\bf How much disk space required?} 10\,MB checked
        out, about 15\,MB after a run, about 400\,MB with the virtualenv,
        about 3\,GB for the Docker image (TeX~Live).
  \item {\bf How much time is needed to prepare workflow?} 5--10 minutes.
  \item {\bf How much time is needed to complete experiments?} less than five minutes.
  \item {\bf Publicly available?} Yes.
  \item {\bf Code licenses?} BSD-3-Clause Clear.
  \item {\bf Data licenses?} CC-BY-4.0.
  \item {\bf Workflow automation framework used?} GNU Make and shell scripts.
  \item {\bf Archived?} \texttt{10.5281/zenodo.21554122}.
\end{itemize}
}

\subsection{Description}

\subsubsection{How to access}

The artifact is archived at Zenodo under the DOI above and mirrored at
\url{https://github.com/huawei-csl/aneto-ae}.

\subsubsection{Hardware dependencies}

None beyond a commodity x86-64 Linux machine.

\subsubsection{Software dependencies}

Python 3.12, GNU Make, \texttt{pdftoppm} from poppler-utils for the figure
comparison in \texttt{verify.py}, and a TeX~Live installation with
\texttt{libertine} and \texttt{newtxmath}, which the figures use to render
their text.
\texttt{scripts/install.sh} creates the virtualenv and installs the
dependencies pinned in \texttt{requirements.txt}. The PROFET implementation compared in
\autoref{sec:exp:profet} is vendored as a Python wheel, so the installation needs
no access to its GitHub repository. The \texttt{Dockerfile} provides the reference environment.

\subsubsection{Data sets}

\texttt{experiments/\allowbreak data/\allowbreak platform.d/\allowbreak <platform>/\allowbreak results/\allowbreak aneto-metrics-v1/}
holds one JSON file per suite, where \texttt{<platform>} is the slug that
\texttt{experiments/config.json} maps to each platform. Each workload contains its measured operating
points across the frequency or latency sweep and names which of them is the
baseline operating point. A point records \texttt{cpi}, \texttt{cycles}, \texttt{instructions},
the average miss penalty \texttt{mp} in core cycles, and misses per
instruction as \texttt{mpi\_fit}. Each hardware point
aggregates at least three repetitions, whereas simulator points are single
deterministic runs. The suites on the hardware platforms are SPEC CPU2017,
GAP, Ligra, GMS, PARSEC, DaCapo, and AI-ML. ChampSim and Sniper run the DPC-4
traces, ChampSim at full \acs{dram} bandwidth with the default prefetchers
enabled (ChampSim-pf) and disabled (ChampSim-nopf).

\subsubsection{Security, privacy, and ethical concerns}

None. The analysis runs offline on the included measurements and uses no
credentials and no personal data. Only the installation needs the network.

\subsection{Installation}

\begin{verbatim}
./scripts/install.sh # venv + pinned deps
# or: docker build -t aneto-ae .
#     docker run --rm -it aneto-ae
\end{verbatim}

\subsection{Experiment workflow}

\begin{verbatim}
./scripts/run_all.sh
.venv/bin/python3 scripts/verify.py
\end{verbatim}

Each experiment's \texttt{main.py} reads the platform JSON files, runs the \model{}
pipeline, and writes its figure, table, or macros to \texttt{out/}. The README maps each experiment to
its figure or table, for example \texttt{cpi\_prediction\_error} to
\autoref{tab:accuracy} and \texttt{memory\_technologies} to
\autoref{fig:tech_comparison}.

\subsection{Evaluation and expected results}

\texttt{scripts/verify.py} exits with zero when every table and quoted number in the
non-ARM scope (see Notes) is reproduced. The \texttt{expected/} directory holds the
reference outputs, one subdirectory per experiment, with the
\texttt{macros.tex} and \texttt{table.tex} that contain the values printed in
the paper, plus the figure PDFs. \texttt{run\_all.sh} writes the reproduced
equivalents to \texttt{experiments/<experiment>/out/}. \texttt{verify.py} then
compares the two directories, macro by macro and table token by token. The pipeline is
deterministic, so values are expected to be string identical. A numeric
difference within a relative tolerance of $10^{-9}$ is reported as SOFT and
still exits with zero, whereas any other difference exits with one. Figure PDFs, and the
\texttt{table.pdf} that \texttt{run\_all.sh} typesets from each
\texttt{table.tex}, are rendered to PNG with
\texttt{pdftoppm} and compared pixel by pixel against the reference copies. A
pixel difference is reported but does not fail the run, because a different
TeX~Live or matplotlib can shift text placement without changing a value. All
eleven experiments match exactly in the Docker reference environment.

\subsection{Experiment customization}

Platforms and the \acs{mp} factor at which the \acs{cpi} error tables are
evaluated are set in \texttt{experiments/config.json}. The \texttt{aneto} package is a small library that runs on
any directory of aneto-metrics-v1 files. Its entry point
\texttt{aneto.\allowbreak run\_leave\_one\_out(dir,\allowbreak{} mp\_factor)}
takes one platform's aneto-metrics-v1 directory and returns the accuracy
metrics, the qualifying workload count, and the per-workload predictions.

\subsection{Notes}

The ARM results come from pre-production vendor hardware
whose measurements we are not permitted to redistribute. The artifact
therefore excludes \autoref{sec:exp:arm}, \autoref{fig:arm_worked}, the ARM
rows of \autoref{tab:platforms}, and the two ARM error percentiles quoted in
the paper's abstract.

\subsection{Methodology}

Submission, reviewing and badging methodology:

\begin{itemize}
  \item \url{https://www.acm.org/publications/policies/artifact-review-and-badging-current}
  \item \url{https://cTuning.org/ae/submission-20201122.html}
  \item \url{https://cTuning.org/ae/reviewing-20201122.html}
\end{itemize}

\end{document}